# Analyzing Heat Transport in Crystalline Polymers in Real and Reciprocal Space

Lukas Reicht[1], Lukas Legenstein[1], Sandro Wieser[1,2], and Egbert Zojer[1,*]

[1] Institute of Solid State Physics, Graz University of Technology, NAWI Graz, Graz, Austria
[2] Institute of Materials Chemistry, TU Wien, Vienna, Austria
* Correspondence: egbert.zojer@tugraz.at

Heat transport can be modelled with a variety of approaches in real space (using molecular dynamics) or in reciprocal space (using the Boltzmann transport equation). Employing two conceptually different approaches of each type, we study heat transport in crystalline polyethylene and polythiophene. We find that consistent results can be obtained when using highly efficient and accurate machine-learned potentials, provided that the physical intricacies of the considered materials and methods are correctly accounted for. For polythiophene this turns out to be comparably straightforward, while for polyethylene we find that the inclusion of higher-order anharmonicities is crucial to avoid a massive overestimation of the thermal conductivity. The responsible long-lived phonons are found at relatively high frequencies between 11 THz and 16 THz. This complicates the use of classical statistics in all molecular-dynamics-based approaches.

## 1 Introduction

Heat-transport properties of crystalline polymers have recently attracted interest due to the possibility of realizing particularly high thermal conductivities in chain direction in highly ordered materials. In the present study, we focus on two prototypical polymeric materials, crystalline polyethylene (PE) as a model for a conventional polymer, and polythiophene (PT), as a reference material for semiconducting polymers. Their crystal structures are shown in Figure 1. Amorphous PE and PT have low thermal conductivities of below 0.5 Wm$^{-1}$K$^{-1}$.[1,2] Contrary, in their crystalline state, much higher thermal conductivities can be achieved along the direction of the polymer chains. This has been demonstrated especially for PE variants, for which ref. [3] summarizes fourteen distinct measurements. Amongst these, the highest thermal conductivities of 104 Wm$^{-1}$K$^{-1}$ [4] and 90 Wm$^{-1}$K$^{-1}$ [5] were obtained for stretched, crystalline PE fibers along the fiber axes. Also, in a variety of thin films, thermal conductivities between 22.5 Wm$^{-1}$K$^{-1}$ and 62 Wm$^{-1}$K$^{-1}$ [2,6–8] were observed parallel to the direction of preferred orientation of the PE chains. Kim et al. even suggest that the "thermal conductivity of PE films has not yet reached its upper limit" and that "the practical challenge is synthesizing disentangled UHMWPE (ultra-high-molecular-weight polyethylene) films with larger extended crystal dimensions".[6] In this context, it would be interesting to know the upper theoretical limit of PE's thermal conductivity for a structurally perfect PE single crystal. Reliably answering this question is one of the goals of the present study. Regarding crystalline PT, there are much fewer data available in literature, but it still serves as an interesting model for conjugated materials with a fundamentally different bonding structure along the backbone (namely, alternating single and double bonds between carbons plus additional, stabilizing carbon-sulfur-carbon bridges). In fact, as will become evident below, an accurate calculation of the thermal conductivity for PT is much more straightforward than for PE, which makes it a well-suited model material. For PT, we are only aware of a theoretical study, which predicts its thermal conductivity in chain direction to be 198 Wm$^{-1}$K$^{-1}$ [9] (which is compared to our results in Supplementary Section S2.1). This value is comparable to the simulated thermal conductivity along PE chains, as can be inferred from a comparison to the corresponding values in Table 1. Considering that all results mentioned above deal with transport along polymer chains and to ease the overall discussion, in the main manuscript we focus solely on heat transport along the chains of the crystalline polymers. The results for transport in the van-der-Waals-bonded directions are provided in Supplementary Sections S3 and S4.

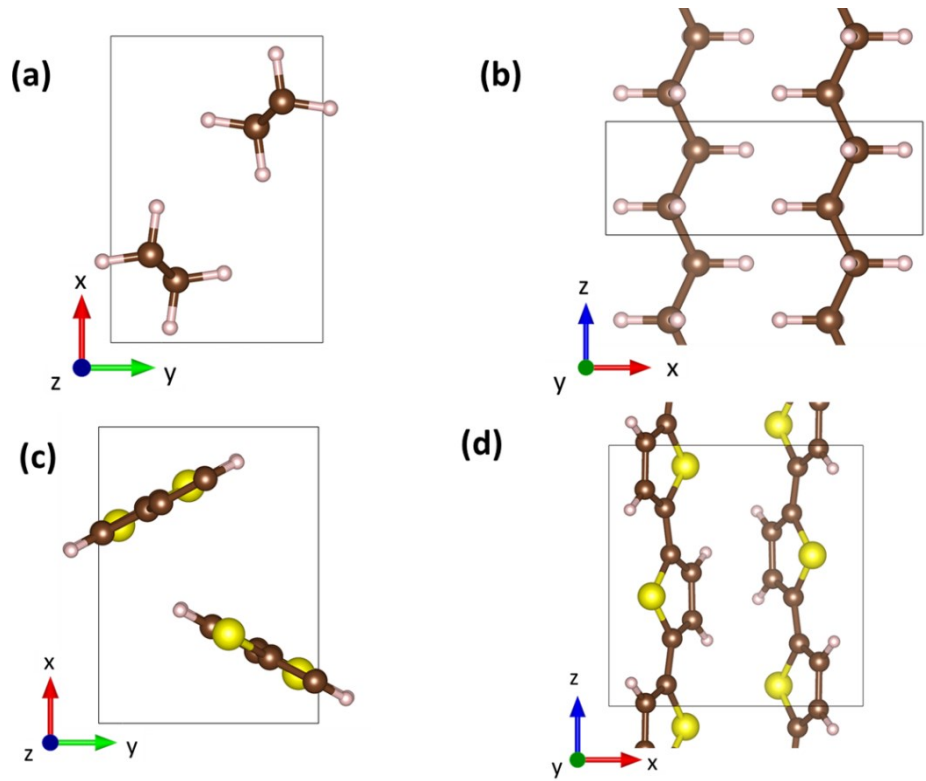

*Figure 1: Crystal structure of (a,b) polyethylene (PE) and (c,d) polythiophene (PT). For both materials, the orthorhombic unit cell (shown with black lines) contains two molecular chains. Carbon, hydrogen, and sulfur atoms are depicted as brown, white, and yellow spheres, respectively. Adapted from ref. [10]; © the authors, published by MDPI under CC-BY 4.0 license.*

A somewhat discomforting observation in Table 1 is the huge variation of calculated thermal conductivities of PE reported in the literature. We attribute this to mainly two aspects: (i) For the description of the interatomic forces, either comparably accurate dispersion-corrected density-functional theory (DFT) approaches or classical, transferable force fields have been used. The latter are, however, notoriously inaccurate, as has recently been shown also explicitly for PE.[10] (ii) Fundamentally different methods for simulating thermal conductivities have been used in different studies. These comprise, on the one hand, reciprocal-space methods, which rely on a description of heat transport caused by the propagation of phonons. These methods typically employ a perturbative approach combining anharmonic lattice dynamics (ALD) and the Boltzmann Transport Equation (BTE).[11] On the other hand, real-space approaches based on molecular dynamics (MD) have been used, like non-equilibrium molecular dynamics (NEMD)[12] and the Green-Kubo (GK)[13,14] method, which is based on employing equilibrium molecular dynamics simulations in combination with the fluctuation-dissipation theorem. One would hope that, even though the aforementioned approaches are fundamentally different, they should yield reasonably similar results. Unfortunately, the vastly different thermal conductivities in Table 1 suggest that this might not necessarily be the case here. Such an observation would be highly problematic, as it would generally cast doubt on the reliability of heat transport simulations. One of the reasons for possible discrepancies are the commonly applied and often inevitable approximations in the MD and BTE simulations: while the former incorporates the materials' full anharmonicity, in the BTE calculations listed in Table 1 only third-order force constants (and, thus, only three-phonon scattering processes) have been considered. Conversely, while in the BTE-based simulations the occupation of phonon modes is described using the (appropriate) Bose-Einstein (BE) statistics, all MD simulations were performed classically, such that the thermal occupation of phonons was described by the equipartition theorem (EQ).[15] This even applies when not explicitly considering phonons in MD simulations, as their properties are still implicitly part of the MD trajectories (especially in crystalline materials). I.e., phonon occupations still form the basis for the real-space thermal motions of the atoms.

*Table 1: Simulated thermal conductivity of polyethylene (PE) along the chain direction, ordered by year of publication. Methods are the Boltzmann transport equation based on anharmonic lattice dynamics (ALD-BTE), non-equilibrium molecular dynamics (NEMD) and Green-Kubo methods (GK). The third-order force constants for the ALD-BTE calculation can be obtained via a finite difference scheme, as it is implemented for example in phono3py[11] or ShengBTE[16], or via the temperature dependent effective potential (TDEP)[17] method. COMPASS[18], AIREBO[19] and REBO[20] are transferrable force fields. In the DFT calculations, Shulumba et al.[21] used the van der Waal density functional by Dion et al.[22], while Wang et al.[23] employed the optB88-vdW[24,25] functional. BE stands for Bose-Einstein statistics, MB for classical Maxwell-Boltzmann, EQ for classical equipartition. To date, to the best of our knowledge, no calculation exists that includes all phonon scattering orders (i.e. full anharmonicity) at DFT accuracy.*

| First Author | Year | Method | Scattering order | DFT functional or force field | Statistics | Thermal conductivity / $Wm^{-1}K^{-1}$ |
|---|---|---|---|---|---|---|
| Zhang[26] | 2020 | ALD-BTE, finite differences | 3 | AIREBO potential | BE | 216 |
| Wang[23] | 2017 | ALD-BTE, finite differences | 3 | optB88-vdW[24,25] functional | BE | 237 |
| Shulumba[21] | 2017 | ALD-BTE, TDEP | 3 | vdW-DF functional[22] | BE<br>MB | 160<br>~255 |
| Zhang[27] | 2014 | NEMD | all | COMPASS | EQ | 50 [§] |
| Henry[28] | 2010 | GK | all | AIREBO potential | EQ | 47 [§] |
| Shen[4] | 2010 | GK | all | AIREBO potential | EQ | 180 ± 65 [§] |
| Ni[29] | 2009 | NEMD | all | REBO potential | EQ | 310 ± 190 |

§ at the time the quoted data were published, the determination of the heat flux was incorrectly implemented for many-body interactions in LAMMPS.[30–32] This likely affects the results, as such interactions are considered in the AIREBO potential and in the COMPASS force field.

In view of the said discrepancies, a central goal of the present paper is to assess what measures need to be taken to provide a quantitatively accurate description of thermal transport in PE and PT that works both in real space (employing MD approaches) and in reciprocal space (employing the BTE). Therefore, the above-mentioned problems need to be assessed and, if possible, overcome. In recent years, the challenge of accurately and efficiently describing interatomic interactions has been largely overcome: system-specific machine-learned potentials (MLPs) have been shown to provide essentially DFT accuracy, while reducing the computational costs by many orders of magnitude.[33,34] In fact, we have recently suggested an approach, in which computationally extremely efficient moment tensor potentials (MTPs) (as implemented in the MLIP code[35]) are parametrized on DFT-calculated reference structures generated during active learning runs using the VASP code.[36–39] This strategy will also be applied here and is described and benchmarked in detail in ref. [40]. There it is also shown that the obtained MTPs allow to fully reproduce the DFT results for elastic and phonon-properties of metal-organic frameworks and that employing the MTPs in conjunction with molecular dynamics approaches yields an amazing quantitative agreement between the measured and calculated single-crystal thermal conductivity of MOF-5. Moreover, in a follow-up study, we illustrated that MTPs achieve DFT accuracy also in calculating phonon-related properties of polyethylene (PE), polythiophene (PT) and poly-3-hexylthiophene (P3HT).[10] Of particular relevance for the present work is the finding that for PE the MTPs achieved essentially DFT accuracy also when calculating the thermal conductivity with the Boltzmann transport equation employing anharmonic lattice dynamics to third order (ALD-BTE).[10] These encouraging results show that the challenge of achieving a highly accurate and at the same time computationally extremely efficient description of interatomic and inter-molecular interactions can nowadays be overcome. Thus, the remaining key challenges are the intricacies of modelling thermal transport per se and doing so both based on phonon properties as well as on real-space particle trajectories.

## 2 Results
### 2.1 Brief overview of approaches for simulating thermal conductivities

Comprehensive and insightful discussions of methodological approaches for simulating heat transport go beyond the scope of the current manuscript and can, for example, be found in refs. [15,41–43]. As discussed above, it is still crucial to address the main aspects of such simulations, as they will be relevant for the later discussion of our results on PT and PE: when aiming at a description of heat transport in reciprocal space via phonons, the most straightforward approach is to use the Boltzmann transport equation (BTE).[11] A linearized form of the BTE can, for example, be solved directly following the approach described by Chaput et al.,[44] which is also implemented in the phono3py code[11] (see Methods section). This direct solution of the linearized BTE is used in the current study and we will refer to it concisely as the "full BTE". A common simplification of the full BTE is realized, when applying the relaxation-time approximation (RTA). In this context, one has to keep in mind that employing the RTA leads to a reduction of the thermal conductivity,[11] an aspect that will become relevant later, when analyzing BTE-calculated thermal conductivities for PE and PT. An advantage of the RTA is that it yields a particularly simple form of the BTE, which can be analyzed in a straightforward manner. It reads[11]

$$\kappa_{RTA} = \frac{1}{N_q V_c} \sum_\lambda C_\lambda \boldsymbol{v}_\lambda \otimes \boldsymbol{v}_\lambda \tau_\lambda^{RTA}, \quad (1)$$

with $\kappa_{RTA}$ representing the thermal conductivity tensor in the RTA and $N_q$ denoting the number of $\boldsymbol{q}$-points used for sampling reciprocal space. The volume of the unit cell is given by $V_c$, the mode heat-capacity by $C_\lambda$, the group velocity vector by $\boldsymbol{v}_\lambda$, and the phonon lifetime by $\tau_\lambda$. The composite index $\lambda$ refers to the wave vector $\boldsymbol{q}$ and the band index $n$. The sum runs over all phonon modes $\lambda$. A particular advantage of this approach compared to, e.g., NEMD and GK is that employing Eq. (1) not only yields a numerical value of the thermal conductivity, but also allows to separately analyze the contribution of each phonon mode. The mode heat capacities $C_\lambda$ and the group velocities $\boldsymbol{v}_\lambda$ can be straightforwardly obtained from harmonic lattice dynamics, employing a finite-differences scheme (see Methods section). More effort is needed to determine phonon lifetimes, for which one has to consider phonon scattering processes between at least three phonons. In these anharmonic lattice dynamics (ALD-BTE) calculations, it is straightforward to apply the Bose-Einstein statistics as the correct quantum statistics for the occupation of the phonon modes. This affects the calculation of mode heat capacities and phonon lifetimes. Alternatively, phonon lifetimes can be obtained from equilibrium molecular dynamics trajectories (MD-BTE). In these simulations, one typically relies on classical statistics (i.e., equipartition statistics in the phonon picture[15]). When aiming at a consistent application of identical statistics in the MD-BTE calculations, one then has the option to apply equipartition statistics in the calculation of the mode heat capacities, when evaluating Eq. (1).

Going beyond the harmonic approximation results in a shift of phonon frequencies due to anharmonic effects.[45] This can change group velocities and heat capacities. This phonon renormalization also leads to a different scattering phase space, i.e., different possibilities for phonon scattering processes that are consistent with energy and momentum conservation. Thus, phonon renormalization can change phonon lifetimes. For heat transport along the chain direction in PE and PT, it, however, turns out that the effects of phonon renormalization are negligible, as shown in Supplementary Section S5. In passing we note that a perturbative approach like the BTE can only be applied as long as the phonons are not overdamped, i.e., as long as their lifetimes are sufficiently large and fulfill the Ioffe-Regel limit[46]. While the Ioffe-Regel limit is not fulfilled for strongly disordered systems,[47,48] it is satisfied for the phonons relevant in PE and PT, as shown in Supplementary Section S6.

Another mechanism that is worthwhile mentioning is the conduction due to wave-like phonon tunneling, which is neglected in the BTE. It has been described by Simoncelli et al. in the framework of the Wigner transport equation.[49] Phonon tunneling gives rise to an additive term that increases the overall thermal conductivity. This is mainly relevant for materials with low thermal conductivity, and we have recently shown that phonon tunneling is crucial for the series of acene crystals.[50] Consistent with this notion, for the polymers considered here, phonon tunneling becomes relevant especially in the vdW-bonded directions, as shown in Supplementary Section S3. For transport along the polymer axes, where particle-like transport is highly efficient, the impact of phonon tunnelling is, however, comparably minor. More quantitatively, for PT, it increases the thermal conductivity along the chains by ca. 2.5%, while for PE the increase is only around 0.2% (see Supplementary Section S3).

When applying anharmonic lattice dynamics for calculating phonon lifetimes, one often describes anharmonicity only to the leading (third-order) term, especially when dealing with materials that are highly complex. In the framework of the BTE that means that only three-phonon scattering processes are taken into account for determining phonon lifetimes. The limitations of this approach will become particularly relevant for PE, making it necessary to discuss alternative approaches. Going beyond the third-order term, one can (at least conceptually) also consider higher orders, such as the fourth-order term in the Taylor expansion of the potential-energy surface. However, until recently, [15,51,52] the resulting four-phonon interactions have rarely been considered in literature, because their calculation is computationally extremely costly. As an example for a recent result, Ravichandran et al. investigated 17 zinc blende structures, where the thermal conductivities of half of them were strongly influenced by four-phonon scattering resulting in an at least 20% reduction of the thermal conductivity at room temperature.[53] Four-phonon scattering becomes particularly important, when three-phonon scattering is reduced by selection rules. This has been observed, for example, for boron arsenide, whose thermal conductivity decreases from 2200 $Wm^{-1}K^{-1}$ to 1400 $Wm^{-1}K^{-1}$ when four-phonon scattering processes are included.[54] Also for graphene four-phonon scattering has been found to be important.[55] Unfortunately, for comparably complex materials like PE and PT, an explicit calculation of four-phonon processes is computationally too demanding. Therefore, we chose to obtain phonon lifetimes from molecular dynamics trajectories using the DynaPhoPy package.[56] In this approach, equilibrium molecular dynamics simulations are performed, during which the velocities of the atoms are tallied. These velocities are then projected onto phonon eigenvectors and onto their associated wave vectors. The autocorrelation of the group velocities is then Fourier-transformed, yielding a power spectrum for each phonon mode. These power spectra consist of a single peak, each of which is fitted by a Lorentzian function from whose width the phonon lifetime is obtained. In this way, the full anharmonicities of the materials are considered, albeit at the price of using classical equipartition statistics. Independent of the statistics used for calculating phonon lifetimes, for the mode heat capacity, one can again choose, e.g., between equipartition and the Bose-Einstein statistics, which leaves a variety of possible combinations. These have been analyzed in detail for selected inorganic crystals by Puligheddu et al.[15]. Here, we will assess the combinations of statistics and degrees of anharmonicity that appear to be most relevant for properly describing heat transport in crystalline polymers.

Besides ALD-BTE and MD-BTE, there is also a multitude of purely MD-based techniques for describing heat transport. A wide-spread approach in this context is the Green-Kubo method,[13,14] where the thermal conductivity is obtained from integrating the time correlation of the heat-flux in an equilibrium MD calculation. As a correction to the implementation of the heat flux for MTPs has only recently become available,[57,58] we do not apply this approach here and, instead, focus on non-equilibrium methods like the reverse NEMD approach.[12] In this method, a constant heat flux between two thermalized regions is generated by repeatedly exchanging the velocities of particles in the "hot" and "cold" regions, which induces a temperature gradient in the material such that the thermal conductivity can then be obtained from the ratio between the heat flux and the temperature gradient. An ambiguity in this context is caused by the fact that in many NEMD studies[59,60] the essentially constant temperature gradient is determined in the "bulk" region far from the hot and cold thermostats, where boundary scattering occurs resulting in a pronounced temperature drop. As an alternative, Li et al.[61] suggested that the net temperature gradient should be determined from the temperature difference and distance of the thermostat regions. In this way also the temperature drops at the region boundaries are included. In the present study, the NEMD results were obtained using the strategy suggested by Li et al. because only this strategy yielded results consistent with AEMD and MD-BTE, as discussed in Supplementary Section S7.

In view of this ambiguity, as an alternative strategy, we also employed the approach-to-equilibrium molecular dynamics (AEMD)[62] method, where the supercell is divided into a hot and a cold region with either a rectangular[62,63] or a sinusoidal[64] temperature profile. In a second step, the system is allowed to equilibrate, and the thermal conductivity of the material is obtained from the time-dependent decay of the (averaged) temperature difference of the two regions. A technical challenge in both NEMD and AEMD is that at the boundary between the different regions additional scattering of phonons occurs. This requires a finite-size extrapolation based on calculations for a series of finite supercells with increasing size. For NEMD simulations, typically a linear relation between the inverse thermal conductivity and the inverse supercell length is assumed. This approach has, however, been cast into doubt (see, e.g., the recent paper by Dong et al.[41]). The finite-size extrapolation for AEMD (with the thermal conductivity scaling linearly with one over the square of the supercell length) has a more rigorous foundation, but is still based on some assumptions such as a linear dispersion relation.[65] Another technical challenge for NEMD and AEMD approaches is that they are computationally significantly more demanding than ALD-BTE calculations. This is particularly relevant for materials with very high thermal conductivities, which require the consideration of extremely large supercells for the finite-size extrapolations. For example, for the case of PE, the largest considered supercell contained more than 100,000 atoms for which the coupled equations of motion were simulated for 4.3 million time steps. Historically, for complex materials like crystalline polymers this required the use of fast empirical potentials like the COMPASS[18] or AIREBO[19] potential (c.f., some entries in Table 1). As these potentials are highly inaccurate, it is not surprising that the four MD-based approaches in Table 1 show a vast spread in thermal conductivities of PE ranging from 47 Wm$^{-1}$K$^{-1}$ [28] to 310 ± 190 Wm$^{-1}$K$^{-1}$. [29]

## 2.2 Expected impact of methodology-inherent approximations on thermal conductivities calculated by reciprocal-space and real-space approaches

To conclude this discussion, it is useful to systematically summarize the impacts of the various methodological aspects, as they are particularly relevant for the remainder of the paper. This is done in Table 2 with the main effects discussed in the following.

Table 2: Influence of different methodological aspects on the heat capacity $C_\lambda$, group velocity $\mathbf{v}_\lambda$, phonon lifetime $\tau_\lambda$ and thermal conductivity κ. The influence is relative to the ideal situation without any approximations. Increases are indicated by ↑, decreases by ↓ and no change by —. Particularly strong increases or decreases are indicated by ↑↑ and ↓↓. For methods that do not explicitly consider $C_\lambda$, $\mathbf{v}_\lambda$, or $\tau_\lambda$, the corresponding columns read "n.a." for "not applicable". For NEMD and AEMD simulations, the impact of using equipartition statistics is expected to be qualitatively similar as when solving the BTE with MD-calculated lifetimes.

| | BTE in the RTA | | | |
|---|---|---|---|---|
| methodological aspect | $C_\lambda$ | $\mathbf{v}_\lambda$ | $\tau_\lambda$ | κ |
| equipartition statistics instead of Bose-Einstein statistics (impacting also NEMD and AEMD simulations) | ↑ | — | ↓ or ↑ | ↓ or ↑ |
| inclusion of fourth-order scattering rate | — | — | ↓ or ↓↓ | ↓ or ↓↓ |
| thermal expansion (assuming positive expansion; opposite trends would apply for negative thermal expansion) | ↑ | ↓ | ↓ or ↑ | typically ↓, but in specific cases ↑ |
| effects beyond the BTE in the RTA | | | | |
| full Boltzmann transport equation (i.e., disregarding the RTA) | — | — | n.a. | usually ↑ |

In the context of BTE-type simulations, using equipartition statistics instead of the Bose-Einstein distribution for calculating phonon occupations results in an overestimation of the mode heat capacity (see Figure 2). This leads also to an overestimation of κ. The deviations are particularly pronounced at high frequencies and at low temperatures. These deviations also affect MD-based approaches like NEMD and AEMD, even though phonons are not considered explicitly. Moreover, when calculating phonon lifetimes from MD trajectories (MD-BTE, see above) the implicit use of equipartition statistics for phonon occupations changes the scattering phase space and, therefore, also the phonon lifetimes, $\tau$.[15] In fact, Puligheddu et al. showed that (at least for MgO and PbTe) this effect dominates over the impact of equipartition statistics on the mode heat capacities.[15] Whether the use of equipartition statistics increases or decreases the phonon lifetimes, can be best described for three-phonon scattering processes: for those, either a phonon decays into two other phonons, or two phonons merge into one phonon. As suggested by Turney et al., when the former process dominates, equipartition leads to higher lifetimes, whereas, when the latter process is dominant, equipartition reduces phonon lifetimes.[66]

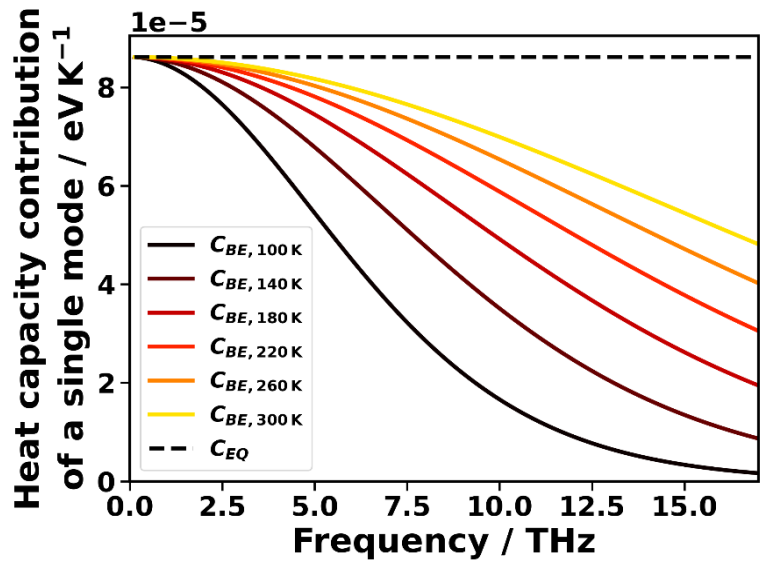

*Figure 2: Heat capacity of a single mode at a given frequency with equipartition statistics $C_{EQ}$ (dotted line) and Bose-Einstein statistics $C_{BE}$ for different temperatures (solid lines).*

Including fourth (and higher) order scattering rates will reduce phonon lifetimes (and thermal conductivities) compared to solely relying on third-order scatterings. In fact, when third-order scattering is particularly low (e.g., due to selection rules), fourth-order scattering can even become dominant and lead to a drastic reduction in phonon lifetimes.[54] Notably, the probability for fourth-order scattering events tends to increase with temperature due to the larger number of available phonons.[53,54] This is opposite to the trend for the difference between equipartition and Bose-Einstein statistics, which drops significantly at higher temperatures. Thus, we suggest that temperature-dependent simulations of the thermal conductivity with both ALD-BTE and MD-BTE can be used to disentangle the role of the two effects, an aspect that will be exploited below.

Another effect that leads to a shift of phonon frequencies is thermal expansion. Since positive thermal expansion generally decreases the bond strength, it results in a downward shift of the phonon bands. The resulting reduced phonon dispersion leads to an increase of the mode heat capacity and a decrease of the phonon group velocities, while the impact on the phonon lifetimes is a priori not certain. Typically, as a net effect, we observed a drop in the thermal conductivity, when considering thermal expansion effects.

The BTE in the RTA typically underestimates the thermal conductivity compared to the full BTE (i.e., the linearized BTE without the RTA).[23,26,67] This underestimation is particularly pronounced for materials with high thermal conductivities,[26] and as such is important for PE and PT in chain direction.

NEMD and AEMD usually employ equipartition statistics, which is expected to cause the same complications as discussed above for MD-BTE type simulations. In contrast, higher-order scattering processes are implicitly accounted for in MD-based simulations and the RTA is not applied. Still, it is worth noting that for MgO and PbTe Puligheddu et al. observed a very good agreement between Green-Kubo based MD results and ALD-BTE simulations.[15] Similar observations were made in other studies.[52,68,69] Relying on an in-depth analysis of the various contributions, Puligheddu et al.[15], however, concluded that the good agreement was at least to some extent the consequence of error cancellation effects arising from opposite impacts of approximations on, e.g., heat capacities and phonon lifetimes and from applying the relaxation time approximation (RTA). Similar error cancellation effects were also suggested by Zhou et al.[70]. Consistent with the error cancellation argument, Puligheddu et al. observed that when calculating phonon lifetimes from MD simulations employing the equipartition theorem, using classical statistics to compute the mode heat capacities yields thermal conductivities that agree more closely with those from ALD-BTE and Green–Kubo MD methods. As a consequence, employing the Bose-Einstein statistics does not serve as a simple quantum-correction for simulations relying on phonon lifetimes obtained using classical MD.

These considerations raise the question, how the situation changes for crystalline polymers, which are structurally much more complex than the inorganic crystals for which the trends were summarized in the previous paragraph. Are consistent results even possible for polymers in view of all mentioned complications? Can one still benefit from cancellation of error effects, or are there additional shortcomings, which further deteriorate the simulations? As the situation turns out to be fundamentally different in PE and PT, these two materials will be discussed separately. This will be done after providing a short description of the machine-learned potentials, which enable meaningful calculations of the types discussed above in the first place. The focus here will be on aspects necessary for understanding the results below, while further technical details can be found in the Methods section.

## 2.3 Interatomic potentials employed in the present study

Performing any of the abovementioned calculations for PT with DFT turned out to be beyond the computational possibilities available on Austria's largest supercomputer[71]. An alternative strategy is the use of highly accurate and efficient machine-learned potentials, which are trained on a few hundred reference configurations of relatively small supercells. In particular, we used moment tensor potentials (MTPs)[35], which are trained on the DFT-calculated reference configurations of an MD-based active learning run[38,39] following a predefined temperature ramp, as implemented in the VASP code[36,37]. The key aspects of this approach are summarized in the Methods section and described in detail in ref. [40]. There, it is also shown that the resulting MTPs are numerically extremely efficient, while at the same time they provide an essentially DFT-quality description of several physical observables, especially in metal-organic frameworks.[40] The same applies to PE and PT, as shown in ref. [10]. In the latter study, we also expanded the strategy for parametrizing optimal potentials for the desired use case: when performing ALD-BTE simulations, phonon properties are calculated from structures containing individual or pairs of slightly displaced atoms. The number of necessary calculations is huge, especially when calculating phonon lifetimes, but still much smaller than when performing MD simulations including finite-size corrections. Therefore, the ideal potentials for ALD simulations, referred to as $MTP^{phonon}$, comprise highly complex variants of MTPs (expressed by a high value of the set level parameter[35]) parametrized on reference structures generated only at low temperatures (up to 100 K). Conversely, for MD simulations, numerically highly efficient (i.e., lower level) MTPs, referred to as $MTP^{MD}$, need to be trained on reference structures containing also large displacements (here, reference structures were generated during active learning runs up to 500 K). As discussed in Ref. [10], the two types of MTPs are not simply "interchangeable"; i.e., it is typically not possible to use an $MTP^{phonon}$ for MD calculations at room temperature, as this will result in a disintegration of the studied materials during the MD run; also $MTP^{MD}$ type potentials are not necessarily ideal for calculating phonon properties as they can potentially yield imaginary phonon bands. All these aspects have been considered for the simulations described in the following. Additionally, to further optimize the performance of the MTPs, we used potentials trained on differently sized unit cells for the 300 K and for the DFT-relaxed structures, as detailed in the Method Section. In passing we note that for the MD-BTE calculations, on the one hand, and for the NEMD and AEMD simulations, on the other hand, ideally the same MTP$^{MD}$s would be used. For technical reasons described in Supplementary Section S11 we had to use not fully consistent potentials for the NEMD and AEMD runs, but this has only a very minor impact on the results, as shown in Supplementary Section S11.

As a final aspect it needs to be stressed that the parametrization of MTPs is a stochastic process. Thus, we always parametrized several (typically five) MTPs, from which the "best" were selected based on their performance in reproducing phonon band structures or in predicting forces occurring in molecular dynamics (see Methods section for details). To get a sense for the spread of the MTPs, in the following not only the data obtained with the "best" MTPs will be given. Additionally, we will report values of certain observables averaged over all MTPs and include the standard deviation of the individual results as a measure for the magnitude of their spread.

## 2.4 Polythiophene: very good agreement between ALD-BTE, MD-BTE, and fully MD-based methods

With accurate MTPs at hand, we perform ALD-BTE calculations of the thermal conductivity of PT along the chain direction at 300 K. This is done using the 300 K unit cell as well as the DFT-optimized cell (corresponding to a temperature of 0 K). The former has been obtained from an MD simulation as described in the Methods section, while the latter is the outcome of a conventional, gradient-based geometry optimization. As the DFT-optimized cells for PT and PE are very similar to MTP-optimized cells (see ref. [10] and Supplementary Section S12), a comparison of thermal conductivities obtained with the 300 K cell and with the DFT-relaxed cell provides insights into the impact of thermal expansion on heat transport. Performing these simulations within the RTA allows one to assess the impact of specific phonons. When employing the "best" MTP$^{phonon}$, the ALD-BTE simulations within the RTA yield rather sizable values of 84 Wm$^{-1}$K$^{-1}$ for the DFT optimized unit cell and a slightly smaller value of 76 Wm$^{-1}$K$^{-1}$ for the 300 K cell. As shown in Table 3, similar values as for the "best" MTPs are obtained when considering the mean values of all five parametrized MTPs. Also the standard deviations for considering all parametrized MTPs are rather small in both cases, which testifies to the consistency of the results.

As discussed above, the RTA usually underestimates the thermal conductivity of materials with high thermal conductivities[23,26] such as PT and PE. Therefore, we also solved the linearized BTE directly,[11,44] referred to as "full ALD-BTE". As expected, the full ALD-BTE yields somewhat higher thermal conductivities of 98 Wm$^{-1}$K$^{-1}$ for the DFT-optimized unit cell and 95 Wm$^{-1}$K$^{-1}$ for the 300 K unit cell. While this approach generally leads to more accurate results, it does not rely on a well-defined relaxation time for each mode, which makes a mode-resolved analysis impossible. Considering that for PT the discrepancy between the RTA and the full ALD-BTE results is not particularly large, it is, thus, useful to analyze the contributions of individual modes to the thermal conductivity on the basis of the RTA simulations.

*Table 3: Thermal conductivity of crystalline PT along the chain direction calculated with different methods. The methods are the linearized Boltzmann transport equation building on anharmonic lattice dynamics (ALD-BTE) in the relaxation time approximation (RTA) and without that approximation (full ALD-BTE), non-equilibrium molecular dynamics (NEMD), approach-to-equilibrium-molecular dynamics (AEMD), and the Boltzmann transport equation with lifetimes from molecular dynamics (MD-BTE). For ALD-BTE calculations, the values in square brackets refer to the standard deviation of the thermal conductivities calculated for five independently parametrized MTP$^{phonon}$s. Uncertainties in NEMD and AEMD results are reported as standard deviations (68% confidence) derived from the finite-size extrapolation (see Supplementary Sections S8 and S10 for details). Heat capacities are treated according to the Bose-Einstein (BE) distribution, $C_{BE}$, or following equipartition statistics (EQ), $C_{EQ}$. In ALD-BTE calculations, lifetimes are treated with Bose-Einstein statistics and the phonon scatterings are taken into account up to third order, while in MD-BTE, NEMD and AEMD calculations, lifetimes follow equipartition statistics and the full anharmonicity is considered. The ratio between the full ALD-BTE and the ALD-BTE in RTA is 91 Wm$^{-1}$K$^{-1}$ divided by 76 Wm$^{-1}$K$^{-1}$. This ratio is used to scale the thermal conductivity of MD-BTE, denoted as MD-BTE($C_{EQ}, \frac{\kappa_{fullBTE}}{\kappa_{RTA}}$). Values from each method that have been made as comparable as possible are marked in bold.*

| Method | Cell | Thermal conductivity / Wm$^{-1}$K$^{-1}$ |
|---|---|---|
| ALD-BTE in RTA | DFT-relaxed | "best": 84<br>mean: 81 [± 5] |
| full ALD-BTE | DFT-relaxed | "best": 98<br>mean: 96 [± 5] |
| ALD-BTE in RTA | 300 K | "best": 76<br>mean: 72 [± 4] |
| full ALD-BTE | 300 K | "best": **91**<br>mean: 86 [± 4] |
| MD-BTE($C_{BE}$) | 300 K | 70 |
| MD-BTE($C_{EQ}$) | 300 K | 83 |
| MD-BTE($C_{EQ}, \frac{\kappa_{fullBTE}}{\kappa_{RTA}}$) | 300 K | **98** |
| NEMD | 300 K | Linear fit: **84** ± 9<br>2$^{nd}$ order fit: 98 ± 12 |
| AEMD | 300 K | **94** ± 3 |

Adding up the thermal conductivity contributions in windows of 1 THz, yields the result shown in Figure 3a by blue bars. Together with the cumulative thermal conductivity shown in Supplementary Figure S3 they illustrate that more than half of the contribution to the thermal conductivity comes from phonons below 4 THz. In fact, it is a common observation that low-frequency phonons are the primary carriers of heat,[72] because of the rapid decrease of phonon lifetimes with frequency. This behavior is indeed found also for PT, as illustrated by the phonon lifetimes superimposed on the phonon band structures in Figure 3b. The largest phonon lifetimes are found for the longitudinal acoustic band at low frequencies and for the two transverse acoustic phonon bands. When considering the mode-resolved contributions to the thermal conductivity in Figure 3c, one, however, sees that the mode contributions of the two transverse acoustic bands are very small. This can be attributed to the very small group velocities of the associated phonons, illustrated by the particularly small slopes of the bands (c.f., Eq. (1)). As a result, the longitudinal acoustic phonons closest to the Γ-point have clearly the largest mode contributions. Nevertheless, as the reciprocal space volume close to the Γ-point is rather small and increases only for larger values of the phonon momentum, **q**, the bin with the largest mode-contribution to the thermal conductivity in Figure 3a is the one between 1 THz and 2 THz. Similar arguments in combination with the often sizable group velocities are responsible for the non-negligible contributions of the bands between 4 THz and 12 THz (see Figure 3a). Notably, they are observed despite the rather small phonon lifetimes and reduced phonon heat capacities (see Figure 2).

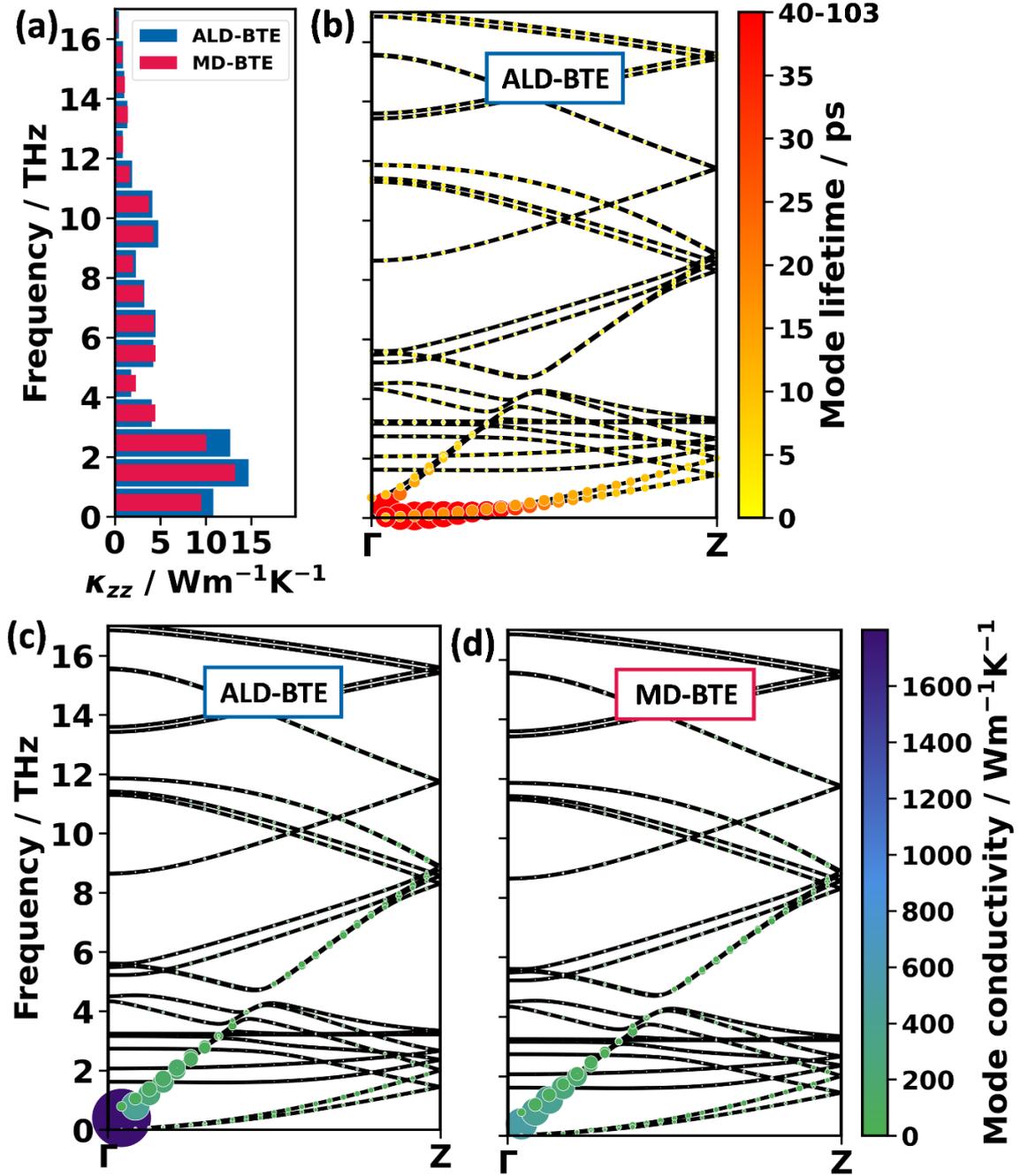

*Figure 3: Panel (a): Histogram of thermal conductivity contributions, $\kappa_{zz}$, of PT in intervals of 1 THz. The bars are calculated as a sum over the entire Brillouin zone, sampled with a 4 × 6 × 48 (ALD-BTE) and 2 × 3 × 48 (MD-BTE) **q**-mesh. Panel (b) shows the lifetimes along the chain direction from the same ALD-BTE calculation. A respective plot for the MD-BTE calculation can be found in Supplementary Section S15. Panels (c) and (d) illustrate the thermal conductivity contribution of the phonon modes for the (c) ALD-BTE and (d) MD-BTE calculation. The areas of the spheres and their colors are scaled linearly with the lifetime and thermal conductivity contribution. Here, the MD-BTE calculation is computed using $C_{BE}$, which gives a qualitatively similar result compared to using $C_{EQ}$ (see Supplementary Section S14). Both calculations are performed for the 300 K unit cell with the MTPs as specified in the Methods section.*

A further advantage of the mode-resolved analysis when applying the RTA is that it allows investigating the role of higher-order scatterings by comparing the ALD-BTE calculation (Figure 3a,c) with MD-BTE results (Figure 3a,d). A possible origin for the somewhat larger deviation of the MD-BTE calculation at the point closest to Γ is discussed in Supplementary Section S13. As argued in Section 2.2, differences between ALD-BTE and MD-BTE primarily arise from whether higher-order phonon scattering effects are considered and also from the use of different phonon occupation statistics. The latter can directly influence the results via the dependence of the mode heat capacities on phonon occupation. Additionally (as described in Section 2.2), the used phonon statistics impact the results also indirectly via the modification of the phonon scattering rates. Nevertheless, for PT in the entire frequency range one observes only minor differences between ALD-BTE and MD-BTE calculated mode contributions to the thermal conductivity, as illustrated in Figure 3a and as seen when comparing Figure 3c and d.

A priori, one cannot exclude that effects due to higher-order scattering and due to the use of classical statistics in MD-BTE simulations might cancel. However, this is very unlikely here, as the impact of employing equipartition instead of the Bose-Einstein statistics should distinctly increase with frequency; i.e., the difference between $C_{EQ}$ and $C_{BE}$ is larger for higher frequencies (see Figure 2 and Supplementary Section S14 for an explicit calculation with both heat capacities). A corresponding increase in the deviations between ALD-BTE and MD-BTE calculated phonon contributions to the thermal conductivity is, however, not observed (see Figure 3a). Thus, based on the data in Figure 3a one can conclude that higher-order phonon scattering processes do not play a significant role in PT. This is consistent with the observation that in PT in the low-frequency region, phonon bands lie densely with no selection rules that strongly inhibit the often dominant third-order scatterings.[53]

As indicated above, when evaluating the BTE with MD-extracted lifetimes (MD-BTE), choosing either $C_{EQ}$ or $C_{BE}$ in the evaluation of Eq. (1) does not qualitatively change the situation (see also Supplementary Section S14). The appeal of using $C_{EQ}$ would be that consistent occupation statistics would be used in the evaluation of the mode heat capacities and when extracting phonon lifetimes from MD trajectories. Conversely, when using $C_{BE}$, at least for the evaluation of the mode heat capacities the physically correct occupation statistics would be used. In practice, also the thermal conductivities are rather similar for both cases, for MD-BTE($C_{BE}$) amounting to 70 Wm$^{-1}$K$^{-1}$ and for MD-BTE($C_{EQ}$) amounting to 83 Wm$^{-1}$K$^{-1}$. Both values agree very well with the ALD-BTE result when also using the RTA (76 Wm$^{-1}$K$^{-1}$). In passing we note that Puligheddu et al. observed a better agreement between MD-BTE and ALD-BTE results when using equipartition statistics for MD-BTE, an aspect they attributed to error cancellations.[15] Such a clear trend is not seen in our simulations. A final aspect relevant in the context of the MD-BTE calculations is that we are not aware of a strategy to seamlessly integrate the MD calculated phonon lifetimes into a full solution of the BTE going beyond the relaxation time approximation. Therefore, we rescaled the MD-BTE($C_{EQ}$) value with the ratio between the aforementioned ALD-BTE results for full BTE simulations and when applying the RTA. This yields a value of 98 Wm$^{-1}$K$^{-1}$ for the rescaled MD-BTE($C_{EQ}$, $\frac{\kappa_{fullBTE}}{\kappa_{RTA}}$) result (see Table 3).

The good agreement between the different simulations discussed so far raises the question whether also purely MD-based approaches would provide consistent results. To address this question, we performed NEMD and AEMD calculations (now employing an MTP as described in Supplementary Section S11). In analogy to MD-BTE, both NEMD and AEMD employ equipartition statistics and capture phonon scatterings to all orders (i.e., they consider the full level of anharmonicity). Additionally, NEMD and AEMD implicitly include phonon renormalization and phonon tunneling effects, even though these effects are small for PT along the chain direction (see above and Supplementary Sections S3 and S5). The NEMD and AEMD results are listed in Table 3. They amount to 84 Wm$^{-1}$K$^{-1}$ ± 9 Wm$^{-1}$K$^{-1}$ (NEMD; linear fit), and 94 Wm$^{-1}$K$^{-1}$ ± 3 Wm$^{-1}$K$^{-1}$ (AEMD). Both values agree excellently with the full ALD-BTE result and with the MD-BTE result rescaled to correct for shortcomings of the RTA (bold values in Table 3).

Regarding the uncertainties of the NEMD and AEMD simulations, three aspects need to be considered: (i) the values in Table 3 are estimated fitting errors from the finite-size extrapolation (see Supplementary Sections S8 and S10) and, consequently, describe the situation for a single MTP$^{MD}$. (ii) To additionally estimate the spread of MTP$^{MD}$s with differently initialized parameters, we performed AEMD simulations on the smallest considered supercells with each of those MTP$^{MD}$s (see Supplementary Section S16). This yielded an uncertainty estimate of about 8%. This value is comparable to (iii) the inherent statistical uncertainty of an AEMD simulation, which we determined by repeating AEMD simulations multiple times with the same MTP, but changing the random seed for initializing the velocities in AEMD simulations (see Supplementary Section S16).

Overall, the excellent agreement between the full BTE result, the rescaled MD-BTE($C_{EQ}$) data, as well as the NEMD and the AEMD calculations is highly encouraging. It shows that consistent results can be obtained independent of the used methodology, as long as all relevant technical peculiarities are considered and as long as the level of anharmonicity included in the different approaches is sufficient. This promising finding shows that it can be worthwhile also for other materials to perform a simultaneous analysis of thermal transport processes in reciprocal space (i.e., via phonon properties and corresponding transport equations) and in real space (i.e., from particle trajectories).

## 2.5 Polyethylene: Why MD-based methods give a lower thermal conductivity than ALD-BTE simulations relying on third-order force constants

The situation becomes considerably more complex for PE despite its seemingly simpler structure: On the one hand, certain technical complications are encountered, which will be discussed first. On the other hand, due to the suppression of three-phonon scattering processes by selection rules, it becomes necessary to go beyond merely calculating third-order force constants when aiming at a consistent description of heat transport in real and reciprocal space.

### 2.5.1 Polyethylene: ALD-BTE type simulations and their technical challenges

Regarding technical aspects, a distinct advantage of PE is that its structure is simple enough that third-order force constants and phonon lifetimes can be calculated using DFT with properly converged numerical settings (see Methods section). This is possible because such simulations for PE require only about 8,000 single-point calculations, whereas for the more complex PT, 50,000 calculations would have been necessary.[10] Consequently, in the following discussion, the MTP results can be directly compared with DFT data. Already in ref. [10], the excellent performance of MTPs had been illustrated for a variety of observables when using the DFT-optimized unit cell (which is essentially equivalent to an MTP-optimized cell – see ref. [10]). This notion is also confirmed by the data in the shaded part of Table 4, which show essentially identical thermal conductivities for the ALD-BTE approach when using DFT or the "best" MTPs. The corresponding values are 296 Wm$^{-1}$K$^{-1}$ (DFT) and 295 Wm$^{-1}$K$^{-1}$ ("best" MTP$^{phonon}$) when applying the RTA and 398 Wm$^{-1}$K$^{-1}$ (DFT) and 408 Wm$^{-1}$K$^{-1}$ ("best" MTP$^{phonon}$) when fully solving the BTE. These data suggest that the thermal conductivity along the chain direction is particularly high and the data show once more that for a high thermal conductivity material like crystalline PE, applying the RTA results in severely underestimated thermal conductivities. Also, when averaging over all parametrized MTPs, similar values of 292 [± 33] Wm$^{-1}$K$^{-1}$ (RTA) and 388 [± 40] Wm$^{-1}$K$^{-1}$ (full BTE) are obtained where the values in square brackets again refer to the standard deviation of the results from five parametrized MTPs. In passing we note that the rather large statistical error is primarily caused by a single MTP that reproduces the general trends but produces a comparably small overall value of the thermal conductivity, as discussed in Supplementary Sections S18 and S19.

When considering the 300 K optimized unit cell, like in PT, the thermal conductivity is reduced. The effect is, however, significantly more pronounced in PE, for which κ drops by between 18% and 29% in the DFT-based ALD-BTE simulations resulting in values of 242 Wm$^{-1}$K$^{-1}$ (RTA) and 309 Wm$^{-1}$K$^{-1}$ (full BTE). This larger drop can be attributed to the more significant thermal expansion of PE compared to PT (for example, the volume changes by 6.6% for PE and only by 1.9% in PT between the DFT-optimized and the 300 K cells).

Table 4: Thermal conductivity of PE along the chain direction calculated with different variants of the ALD-BTE method and different unit cells. DFT-calculated thermal conductivities are denoted by $\kappa^{DFT}$, while MTP-calculated values are denoted by $\kappa^{MTP}$. Five independent MTPs are parametrized, of which we report the "best" (see Methods section for definition of "best"), the mean, and the standard deviation (in square brackets). The DFT values with the 300 K unit cell are used in the subsequent comparison to the MD-based approaches and are therefore highlighted in bold.

| Method | Unit cell | $\kappa^{DFT}$ / Wm$^{-1}$K$^{-1}$ | $\kappa^{MTP}$ / Wm$^{-1}$K$^{-1}$ |
| --- | --- | --- | --- |
| ALD-BTE, RTA | DFT-relaxed | 296 | "best": 295 (mean: 284 [± 34]) |
| ALD-BTE, full | DFT-relaxed | 398 | "best": 408 (mean: 388 [± 40]) |
| ALD-BTE, RTA | 300 K | **242** | "best": 263 (mean: 324 [± 41]) |
| ALD-BTE, full | 300 K | **309** | "best": 393 (mean: not meaningful§) |

§ One of the five MTPs yielded a clearly incorrect thermal conductivity, rendering the calculation of a mean value not meaningful, as discussed in more detail in Supplementary Section S18.

Also, in the calculations performed with the "best" MTP$^{phonon}$ a drop of $\kappa$ to 263 Wm$^{-1}$K$^{-1}$ (in the RTA) and to 393 Wm$^{-1}$K$^{-1}$ (for the full BTE) is observed. This drop is, however, much less pronounced than in the DFT simulations; even more problematic is the observation that this trend is not obeyed for all MTPs, such that the mean value for the five parametrized MTPs even increases to 324 Wm$^{-1}$K$^{-1}$, when applying the RTA. Even worse, when solving the full BTE for all tested **q**-meshes one finds an MTP that produces an unrealistically high value for the thermal conductivity, which in this case prevents an averaging altogether (see Supplementary Section S18). The origin of these problems with the MTPs for the 300 K unit cell of PE remains elusive as they do not occur in any other situation we have ever encountered (PT above, PE for the DFT-optimized cell, molecular crystals in ref. [50], and metal-organic frameworks in ref. [40] and as unpublished data). The problems also prevail despite numerous tests (including the repeated generation of training data, varying the displacement amplitude in phono3py between 0.01 Å and 0.11 Å, and varying the **q**-mesh). In this context it is also worth noting that no such inconsistencies occur in the MD-BTE simulations, where the thermal conductivities calculated with the five independently parametrized MTP$^{MD}$s for the 300 K unit cell display a standard deviation of only 7%. These five MD-BTE calculations also give qualitatively similar mode resolved contributions (see Supplementary Section S17). Likewise, for the AEMD simulations, the MTP$^{MD}$s display a benign behavior with a standard deviation of 5% between the five independent MTP$^{MD}$s (see Supplementary Section S16). In view of these observations, we attribute the complications in the ALD-BTE simulations of PE in the 300 K unit cell to numerical instabilities in the full ALD-BTE calculation, which is apparently rather sensitive to changes in the third-order force constants.

As a consequence, in all following comparisons, we will consider the DFT-calculated data whenever referring to the ALD-BTE results for PE. As already mentioned above, the complications discussed in the current paragraph are not relevant for any of the MD-based approaches discussed next, which is good news, as none of these calculations could be reasonably done using DFT.

### 2.5.2 Polyethylene: including MD-based approaches, which go beyond three-phonon scattering processes

Notably, the aforementioned technical complications are not the only unusual observation in which PE fundamentally differs from PT. Even more importantly, when comparing the ALD-BTE to the MD-BTE results (employing the Bose-Einstein statistics for the mode heat capacities), the thermal conductivities drop by a factor of around three to 95 Wm$^{-1}$K$^{-1}$ for the DFT-optimized unit cell and to 78 Wm$^{-1}$K$^{-1}$ for the 300 K cell. The drop is somewhat less pronounced (but still dramatic) when employing the equipartition theorem to 135 Wm$^{-1}$K$^{-1}$ and 114 Wm$^{-1}$K$^{-1}$, respectively.

This raises the question whether the ALD-BTE or the MD-BTE values are more appropriate. Additionally, it is a priori not obvious, whether it is the inclusion of higher-order phonon scattering or the use of different statistics for describing phonon occupations (or both) that cause the massive differences between ALD-BTE and MD-BTE results in PE, while it had no impact in PT.

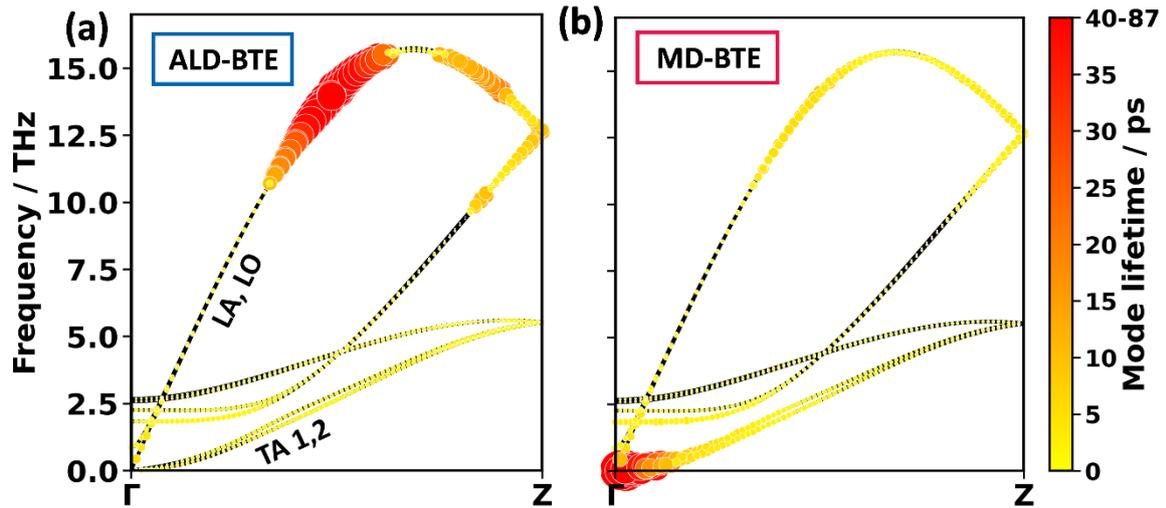

*Figure 4: Phonon band structure of PE along the chain direction with superimposed spheres, whose areas and colors are scaled linearly with the phonon lifetime. The displayed data have been calculated for the 300 K unit cell. Panel (a) shows the ALD-BTE calculation performed with DFT. Panel (b) is the MD-BTE calculation performed with the MTP. The longitudinal acoustic (LA), longitudinal optical (LO), and the two transverse acoustic (TA) bands are indicated. In passing we note that for the MD-BTE simulations the TA modes close to the $\Gamma$−point display particularly large phonon lifetimes. As shown in Figure 5, these modes are, however, not relevant for the thermal conductivities, as they display very small phonon group velocities.*

To address these questions, it is useful to first discuss the phonon lifetimes of PE. The values calculated from the third-order force constants (c.f., ALD-BTE) are superimposed on the phonon band-structures for wavevectors parallel to the directions of the polymer chains in Figure 4a. The equivalent plot for phonon lifetimes obtained from MD trajectories (c.f., MD-BTE) is provided in Figure 4b. Surprisingly, when considering only three-phonon scattering processes (Figure 4a) by far the highest lifetimes are obtained in the frequency range between 11 THz and 16 THz rather than for the low-frequency phonon, for which lifetimes are more than an order of magnitude smaller. This atypical behavior has already been observed by Wang et al.[23] for a single PE chain. They explained it by certain (usually dominant) three-phonon scattering channels not being available for the said phonons: Typically, the dominant scattering mechanism affecting longitudinal acoustic modes in PE is Umklapp scattering involving the twisting and transverse modes, e.g., a longitudinal acoustic phonon (LA) decaying into a twisting acoustic phonon and a transverse acoustic phonon from branch 1 (TA1).[23] This channel is very efficient for LA modes below 11 THz. However, above 11 THz the LA mode has more than twice the energy of the other two modes such that this channel becomes energetically forbidden for three-phonon scattering processes. This leads to the extreme increase in phonon lifetimes around 11 THz to 16 THz. As the corresponding phonons are also characterized by rather high group velocities and appreciable mode heat-capacities, also their mode contributions to the thermal conductivity dominate. This is illustrated in Figure 5a by the blue bars for the integral mode contributions in the 11-16 THz windows and for the phonons along the $\Gamma$−Z-path in Figure 5b. The cumulative thermal conductivity plotted in Supplementary Figure S4 reveals that the modes between 11 THz and 16 THz contribute 79% of the thermal conductivity for the 300 K unit cell. A similar behavior is observed for the DFT-optimized cell, as shown in ref. [10]

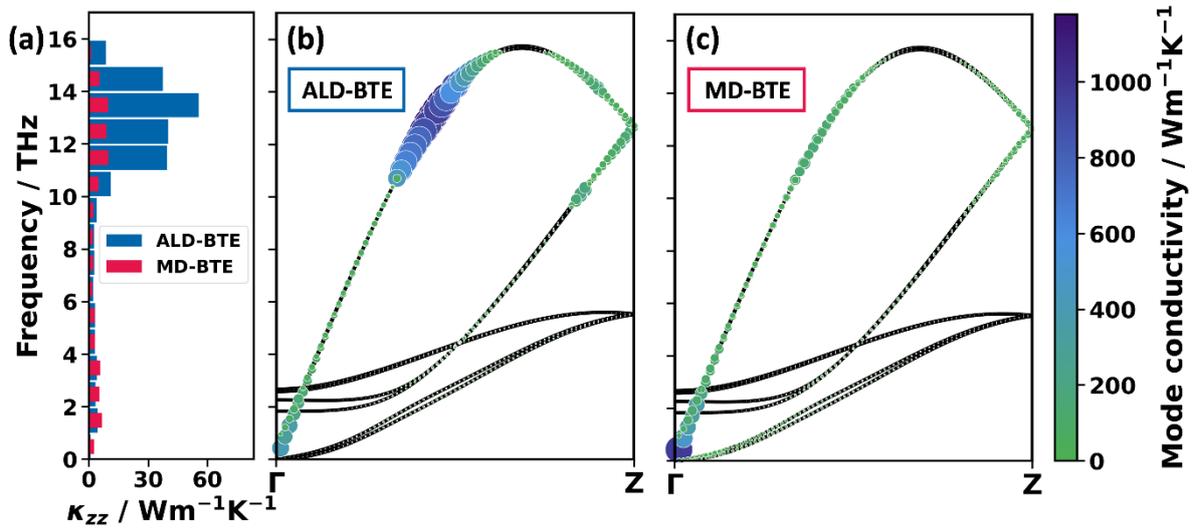

*Figure 5: Panel (a): Histogram of thermal conductivity contributions, $\kappa_{zz}$, of PE in intervals of 1 THz. The bars are calculated as a sum over the entire Brillouin zone, sampled with a 4 × 6 × 160 (ALD-BTE) and 2 × 3 × 160 (MD-BTE) **q**-mesh. Panels (b) and (c) give the thermal conductivity contribution of the phonon modes for the (c) ALD-BTE and (d) MD-BTE calculation. The areas of the spheres and their colors are scaled linearly with the lifetime and thermal conductivity contribution. Smaller spheres are plotted in front such that they are not occluded. Calculations are performed with the Bose-Einstein heat capacity and the 300 K unit cell.*

In this context, it is critical to realize that the above discussed energy-derived selection rules exclusively apply to three-phonon scattering processes. That means that phonons at 11-16 THz can still scatter with phonons below 5.5 THz, when one considers four- or higher-order phonon scatterings. Such processes are, however, neglected in the ALD-BTE approach implemented here. This runs the risk of massively overestimating contributions from 11 THz to 16 THz phonons. That this is actually the case can be inferred from comparing the mode-resolved data for the ALD-BTE and for the MD-BTE calculations in Figure 5. When including full anharmonicity (i.e., in the MD-BTE case) significantly reduced phonon lifetimes and mode-contributions to the thermal conductivity are obtained in the range between 11 THz and 16 THz. Such reductions in thermal conductivity have also been observed for other materials with high thermal conductivity, such as silicon and diamond.[54] Nevertheless, even in the MD-BTE calculations the contributions at 11-16 THz are far from negligible, which makes sense, as the absence of three-phonon scattering processes due to selection rules still increases phonon lifetimes, just not to the extent predicted by the ALD-BTE calculations.

A direct consequence of high-frequency phonons still contributing significantly also in the MD-BTE simulations is that for PE there are comparably large differences in the thermal conductivities when using different statistics for describing the mode heat capacities (see Table 5). This is because for a given temperature, the differences in mode heat capacities increase with increasing frequency (see Figure 2). This explains why switching from $C_{BE}$ to $C_{EQ}$ in PT results only in a moderate increase in the thermal conductivity of 19% (see Table 3), while the increase is rather significant in PE amounting to 46% (see Table 5).

A complication in assessing the role of higher-order scattering processes is that their impact in the MD-BTE calculations cannot be straightforwardly disentangled from effects that result from employing the equipartition theorem. In fact, we hold the use of classical statistics for the determination of phonon lifetimes in the MD-BTE calculations responsible for the increased mode contributions between 0 THz and 4 THz that is apparent in Figure 5a. This assessment is based on the observation that the differences between low-frequency ALD-BTE and MD-BTE mode contributions diminish at higher temperatures, at which the classical equipartition statistics and the Bose-Einstein statistics converge (see Figure 6). In contrast, the huge (relative) difference between ALD-BTE and MD-BTE calculations between 11 THz and 16 THz even increases at higher temperatures. This would be at variance with occupation statistics being responsible for the different phonon lifetimes between 11 THz and 16 THz and provides a key argument for the differences being caused by higher-order scattering events.

Table 5: Thermal conductivity of crystalline PE along the chain direction calculated with different methods. Uncertainties in NEMD and AEMD are reported as standard deviations (68% confidence) derived from the finite-size extrapolation (see Supplementary Sections S8 and S10 for details). Heat capacities are treated according to the Bose-Einstein (BE) distribution, $C_{BE}$, or following equipartition statistics (EQ), $C_{EQ}$. The ratio between the full ALD-BTE and the ALD-BTE in RTA is 309 Wm$^{-1}$K$^{-1}$ divided by 242 Wm$^{-1}$K$^{-1}$ (see Table 4). This ratio is used to scale the thermal conductivity of MD-BTE, denoted as MD-BTE($C_{EQ}, \frac{\kappa_{fullBTE}}{\kappa_{RTA}}$).

| Method | Unit cell | Thermal conductivity / Wm$^{-1}$K$^{-1}$ |
|---|---|---|
| MD-BTE($C_{BE}$) | DFT-relaxed | 95 |
| MD-BTE($C_{EQ}$) | DFT-relaxed | 135 |
| MD-BTE($C_{BE}$) | 300 K | 78 |
| MD-BTE($C_{EQ}$) | 300 K | 114 |
| MD-BTE($C_{EQ}, \frac{\kappa_{fullBTE}}{\kappa_{RTA}}$) | 300 K | 146 |
| NEMD | 300 K | linear fit: 115 ± 20<br>2$^{nd}$ order fit: 146 ± 22 |
| AEMD | 300 K | 127 ± 3 |

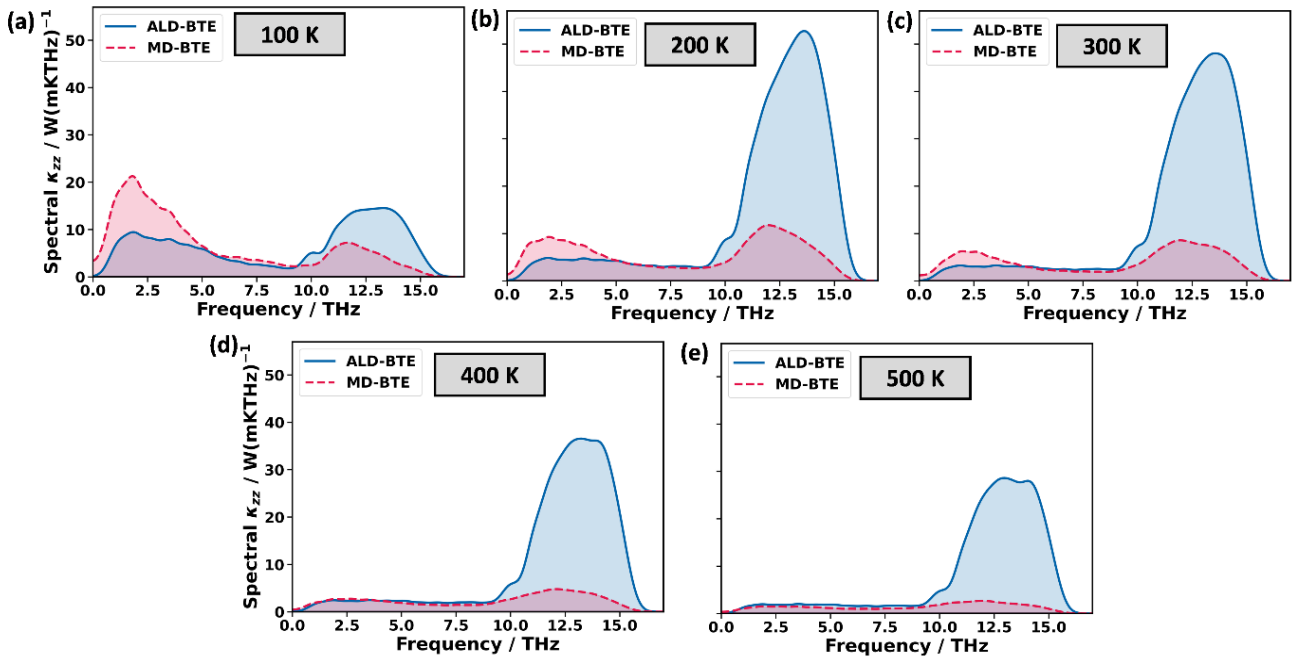

Figure 6: Spectrally resolved contributions to the thermal conductivities of ALD-BTE and MD-BTE at temperatures ranging from 100 K to 500 K. In passing we note that PE melts at around 400 K, but for the timescales on the order of ns that are employed in the MD simulations, this does not have a major impact on the simulations. The spectrally resolved contributions to the thermal conductivity are calculated as the derivative of the cumulative thermal conductivity. The cumulative thermal conductivity is smoothed by a 1-D Gaussian filter with 1 THz standard deviation for the Gaussian kernel; the spectrally resolved contributions to the thermal conductivity are smoothed with a 4 THz standard deviation.

Like for PT, we also performed NEMD and AEMD simulations for PE. As described already for PT, to compare the MD-BTE to NEMD and AEMD, it is useful to consider the mode heat capacity calculated with equipartition statistics and to rescale the MD-BTE results with the ratio between the full ALD-BTE calculation and the ALD-BTE calculation in the RTA. This yields a thermal conductivity of 146 Wm$^{-1}$K$^{-1}$ for the rescaled MD-BTE thermal conductivity (see line labeled as MD-BTE($C_{EQ}, \frac{\kappa_{full\,BTE}}{\kappa_{RTA}}$) in Table 5). Notably, this value compares very well with that of the finite-size extrapolated AEMD calculations, which amounts to 127 Wm$^{-1}$K$^{-1}$. It is also close to the outcome of the NEMD calculations, which depending on the extrapolation scheme yield 115 Wm$^{-1}$K$^{-1}$ for the linear fit and 146 Wm$^{-1}$K$^{-1}$ for the 2$^{nd}$ order polynomial fit (see discussion in Supplementary Section S8). The rather pronounced difference between the two different extrapolation schemes can be attributed to the material's very high thermal conductivity and the associated large phonon mean free paths (of up to 1.5 μm for the ALD-BTE calculation). As a consequence, even the extremely large considered maximum unit cells of PE containing 1440 repeat units in chain direction and in total ca. 103,000 atoms (system sizes only possible due to the extreme numerical efficiency of MTPs) are still too small for performing fully reliable finite-size extrapolations. This is discussed in more detail in Supplementary Section S8.

## 3 Discussion

When describing thermal transport in complex materials like crystalline polymers, one has to overcome two challenges: (i) describing interatomic interactions in a numerically efficient way at a high level of accuracy, and (ii) choosing a suitable theoretical approach for modelling transport processes either in reciprocal space, i.e., based on phonon properties, or in real space, i.e., based on atom trajectories. The first challenge can nowadays be overcome using machine-learned potentials,[10,33,34,40] which we also apply here. In particular, we use moment tensor potentials[35], whose parametrizations are system specific and tailored to the particular use case.[10] This provides essentially DFT-level accuracy combined with a speedup of many orders of magnitude compared to DFT especially when studying large supercells. When describing heat transport via phonons employing the Boltzmann transport equation (BTE), the key challenge is to obtain phonon lifetimes, which we obtain here either from anharmonic lattice dynamics (ALD-BTE) or from the trajectories of molecular dynamics runs (MD-BTE). ALD-BTE is restricted to three-phonon scattering processes (since the direct computation of four-phonon scattering processes is prohibitively expensive for systems as complex as the ones studied here). An advantage of ALD-BTE is that it applies the correct quantum-statistics for phonon occupations. Conversely, MD-BTE considers the full anharmonicity of the potential energy surface but describes phonon occupations only via the equipartition theorem. Despite these fundamental differences, we observe for PT an excellent agreement between ALD-BTE and MD-BTE not only regarding the overall thermal conductivity but also concerning the contributions of individual phonon modes. Importantly, for PT, the results of these two reciprocal space approaches also agree very well with those obtained when describing heat transport in real space either with non-equilibrium molecular dynamics (NEMD) or approach-to-equilibrium molecular dynamics (AEMD) simulations. In this context it is worth mentioning that the excellent agreement between the various, fundamentally different approaches is obtained even though NEMD and AEMD simulations rely solely on classical statistics. This good agreement can be explained by the fact that the main heat carriers are phonons with low frequencies (below about 4 THz), and that the classical equipartition statistics and the Bose-Einstein statistics converge towards each other at low frequencies. Based on these arguments and the excellent agreement between ALD-BTE, MD-BTE, NEMD and AEMD, we conclude that the respective approximations of these methods—namely neglecting four-phonon scattering in ALD-BTE and employing classical statistics in MD-BTE, NEMD and AEMD—have only a minimal impact in PT.

The situation changes fundamentally for PE, where due to the simpler structure of the polymer chain and, thus, due to the reduced number of phonon bands, three-phonon scattering processes in certain frequency regions are forbidden by energy-based selection rules. This results in particularly high phonon lifetimes for the states between 11 THz and 16 THz such that they dominate thermal conduction in the ALD-BTE simulations. The large number of high lifetime phonons with appreciable group velocities results in extremely high thermal conductivities of 309 Wm$^{-1}$K$^{-1}$ at 300 K. A comparison to the MD-BTE simulations, however, suggests that this is an artefact of neglecting higher-order scattering phenomena, whose inclusion massively reduces the phonon lifetimes at high frequencies and, thus, reduces the thermal conductivity by more than a factor of two. Nevertheless, the phonons between 11 THz and 16 THz are still comparably long-lived also in the MD-BTE simulations as they are still not subject to the usually dominant three-phonon scattering. Thus, even in the MD-BTE simulations, their contribution somewhat exceeds that of the lower-frequency phonons. This leads to a complication because for phonons with high frequencies, Bose-Einstein statistics and equipartition statistics differ substantially. Indeed, for PE this results in a sizable difference when comparing MD-BTE simulations using the two statistics for describing mode heat capacities. This suggests that when higher-frequency phonons also provide significant contributions to the thermal conductivity, as it is the case in PE, the impact of using classical statistics can become sizeable.

The main shortcoming of the ALD-BTE simulations of PE, as performed here (and, to the best of our knowledge, in all existing literature on ALD-BTE simulations of PE [21,23,26]) is the neglect of higher-order scattering processes. That higher-order scatterings are important can be inferred, for example, from the observation that the differences between ALD-BTE and MD-BTE simulations in the high-frequency region increase with temperature. This must be a consequence of the dominance of higher-order scattering processes, as the use of different statistics would diminish with temperature.

Overall, by using an accurate, yet numerically extremely efficient description of inter-atomic interactions, we are able to show that describing heat transport either in real space (via particle trajectories in MD approaches like NEMD or AEMD) or in reciprocal space (via phonons and their lifetimes employing the BTE) can produce consistent results for the thermal conductivities also of rather complex materials like crystalline polymers. However, this requires that the simulations accurately capture the physical intricacies of the materials. This can become rather challenging for an "ill-behaved" material like polyethylene, where high-frequency phonons (between 11 THz and 16 THz) significantly contribute to the thermal conductivity. This requires an accurate description of higher-order phonon processes and at the same time challenges the use of classical statistics prevalent in MD approaches. A particular irony in this context is that the challenges encountered in PE are a direct consequence of the comparably less complex structure of the material, which results in selection rules stemming from energy conservation.

# 4  Methods

For the current manuscript, we mostly employed the same settings as in our previous publication, where we thoroughly benchmarked the accuracy of MTPs for describing crystalline polymers.[10] There, we demonstrated different ways to improve the accuracy of the MTPs and we showed the (compared to DFT) very encouraging performance, when calculating phonon related properties, like the phonon band structure, the thermal conductivity via the ALD-BTE approach, thermal expansion, and elastic constants.

## 4.1 Density functional theory and active learning

VASP (Version 6.3.0[37]) was employed for two different tasks: VASP was used to perform ALD-BTE calculations of PE using the phono3py package[11]. Additionally, we used VASP to generate training data for PE and PT during active learning MD runs.[38,39] For the active learning runs, temperature ramps from 15 K to 100 K and from 15 K to 500 K were employed (see Section 4.2 for details). Simulations were performed in an *NpT* ensemble and simulations from 15 K to 100 K were additionally performed in an *NVT* ensemble with the unit cell fixed to the 300 K unit cells (as also explained further in Section 4.2). In every case, a time step of 0.5 fs was used, and 150 MD steps were performed per Kelvin temperature increase. This yielded 72750 MD steps for the simulation from 15 K to 500 K. In these simulations, a Langevin thermostat with a friction coefficient of 10 ps$^{-1}$ was used and the fictitious mass of the lattice degrees-of-freedom was set to 1000 amu. For the training data, the energy cutoff was set to 900 eV (PE) and 700 eV (PT). Convergence tests for these cutoffs and for the k-point sampling can be found in the Supplementary Materials of ref. [10]. For the ALD-BTE calculation of PE with DFT, due to the high computational cost, we initially employed a 700 eV cutoff. For the sake of comparison, we also performed an ALD-BTE calculation for one of the PE unit cells (the one optimized at 300 K) with a cutoff of 900 eV. This yielded a somewhat higher thermal conductivity of 309 Wm$^{-1}$K$^{-1}$ (instead of 288 Wm$^{-1}$K$^{-1}$ for the 700 eV cutoff). This has no impact on the above discussion and is in the range of "benign" variations of clearly below 10%, which we typically observe when changing details of the computational settings (e.g., when switching between MTPs in MD-based simulations as described in Supplementary Sections S16 and S17). Thus, the calculation with the DFT-relaxed cell was performed with a 700 eV energy cutoff and the one for the 300 K cell with a 900 eV cutoff. For the DFT simulations, we employed the projector augmented wave method[36,73] and the PBE functional[74] complemented by Grimme's DFT-D3 correction[75,76] with Becke-Johnson damping.[77] This functional and vdW-correction has been shown to accurately describe phonons of organic semiconductors.[78–81] The Gaussian smearing of the occupancies was set to 0.05 eV.

## 4.2 Moment tensor potentials

We used moment tensor potentials (MTPs) as extremely efficient system-specifically parametrized machine-learned potentials. For a detailed description of their nature, we refer to the literature.[35,82] In a previous publication we benchmarked MTPs and devised a protocol for their efficient use in combination with VASP-based active learning for metal-organic frameworks.[40] This strategy was then extended to polymers and to the specific needs of the calculation of higher-order force constants.[10] Thus, here only a short description of the procedure is provided. DFT-calculated training data were generated in VASP during an active learning MD run, as described in the previous section. MTPs were trained using the MLIP code[35] (version 2). Training of the MTP stopped when the normalized cost function (built from weighted deviations in forces, energies and stresses between MTP results and reference data) dropped by less than a factor of $10^{-3}$ over the previous 50 iterations. This is the default setting in MLIP. For the calculation of the cost function, we used the default weights of 1 (eV)$^{-1}$, 0.01 Å(eV)$^{-1}$, and 0.001 (GPa)$^{-1}$ for energies, forces, and stresses, respectively. The radial basis set size of the MTPs was set to 10. Carbon atoms of PT were split into two atom types, namely those that are bonded to sulfur and those that are not. This drastically improved the accuracy of the MTPs.[10]

In our previous work[10], we showed that it is highly advantageous to consider lattice dynamics and molecular dynamics as two distinct use cases and to choose the hyperparameters accordingly such that optimal performance can be achieved in each of the two cases. For the MTPs used in MD simulations, we, thus, considered the full set of training data obtained in active learning runs from 15 K to 500 K in the *NpT* ensemble (see Section 4.1). The resulting MTPs are referred to as MTP$^{MD}$. For the MTPs used in the BTE simulations, we only considered the subset of training structures generated between 15 K and 100 K, which improves the accuracy of MTPs in describing the phonon band structure by around a factor of two as compared to the case when 15-500 K data are used. This improvement arises because when using only low-temperature training data, the atomic displacements are rather small and are similar to the displacement encountered in phonon band structure or BTE calculations.[10] The MTP trained only on the low-temperature data is referred to as MTP$^{phonon}$. Another factor to consider is the level of the MTP, which determines the size of the basis set and consequently the number of parameters.[35,82] MTPs with higher level are somewhat more accurate, but clearly slower. Since the calculation of third-order force constants for the BTE calculations is very efficient with MTPs, we could use a higher level of 26 for MTP$^{phonon}$. For MD calculations, where computational speed is more important, we had to use a lower level. Thus, the MTP$^{MD}$s are level 22 potentials. A further differentiation is that the MTP$^{phonon}$s are either applied to the DFT-relaxed unit cell or the 300 K unit cell. When performing MD simulations in an *NpT* ensemble at 15 K to 100 K, the unit cell is similar to the DFT-relaxed unit cell and thus the resulting MTPs are well suited for describing the DFT-relaxed unit cell. However, the 300 K unit cell is substantially different from the DFT-relaxed unit cell (especially for PE). Therefore, we created a separate training set sampled at 15 K to 100 K in an *NVT* ensemble with the unit cell fixed to the 300 K unit cell for parametrizing MTP$^{phonon}$s for that cell.

The performance of the five MTP$^{phonon}$s, which were independently parameterized for each set of training data and unit cells with different random initializations, were then ranked via the root mean square deviation (RMSD) between the MTP-calculated and the DFT-calculated phonon frequencies in the entire Brillouin zone up to 12.5 THz (where the exact value of that cutoff has virtually no impact on the ranking, as shown in ref. [10]). The performance of MTP$^{MD}$s was evaluated on structures sampled at 300 K in a VASP active learning run, which was produced independent of the generation of the training data. Here, the MTP$^{MD}$ with the smallest RMSD for the calculated forces on atoms is chosen as the "best" one.

The NEMD and AEMD simulations (and here especially the finite-size extrapolation) generated gigantic computational efforts caused by the very large thermal conductivity along the chain direction. This required huge supercells in chain direction with up to 103,680 atoms to at least approach the mean free path of the phonons and the need of solving the equations of motion 4.3 million times. At the point in time the simulations were performed, the optimized procedure for parametrizing MTPs outlined in ref.[10] was not yet fully established and we used a somewhat different (first-generation) parametrization strategy for the MTPs as described in Supplementary Section S11. To make sure that this does not result in serious deviations, we repeated the AEMD simulation with consistent MTP$^{MD}$s for individual, relatively small unit cells. For the considered unit cell lengths (three per material) deviations in the same range as those found between AEMD thermal conductivities calculated with independently parametrized MTP$^{MD}$s were obtained (see Supplementary Sections S11 and S16). This suggests that the first-generation MTPs are acceptable for the analysis provided in the current paper.

## 4.3 Unit cell at 300 K from molecular dynamics

All MD calculations were performed in LAMMPS[83] (Large-scale Atomic/Molecular Massively Parallel Simulator) using the version from the 2$^{nd}$ of July 2021, whereby we used the LAMMPS-MLIP interface (version 2)[35]. For the NEMD, AEMD, MD-BTE and ALD-BTE calculation, we chose a unit cell that corresponds to the average lattice parameters at 300 K. To find this thermally expanded unit cell, we performed MD runs with the first-generation MTPs described in the previous section. These MTPs yield unit cells very well consistent with those obtained with MTP$^{MD}$s, which are provided in in ref. [10]. In the MD runs, we used a 4 × 6 × 12 supercell for PE and a 4 × 6 × 4 supercell for PT with extents of approximately 30 Å in each direction. The time step was set to 0.5 fs. The converged simulation times were 500 ps for PT and 700 ps for PE. The first 50 ps (PT) and 100 ps (PE) were neglected in the averaging to account for equilibration effects. A Langevin thermostat with a 0.1 ps damping coefficient was used in these simulations.

## 4.4 Non-equilibrium molecular dynamics

For the non-equilibrium molecular dynamics (NEMD) simulations[12], the unit cell was fixed to the one that was obtained from the thermal expansion calculation at 300 K (see above). While we used the Langevin thermostat for the initialization of the NEMD calculation, in the actual NEMD run, we used the Müller-Plathe method in an *NVE* ensemble.[12] The temperature gradient was evaluated as described by Li et al. by taking the difference between the temperatures of the hot and cold region.[61] There exists an alternative definition of the temperature gradient, which is discussed in Supplementary Section S7. We, however, note that only when defining the temperature gradient as suggested by Li et al.[61], results consistent with AEMD and MD-BTE simulations were obtained. The finite-size extrapolation as well as further details on the NEMD simulation are provided in Supplementary Sections S7 and S8.

## 4.5 Approach-to-equilibrium molecular dynamics

In approach-to-equilibrium molecular dynamics (AEMD)[62,63], one half of the simulation box is heated to a hot temperature $T_1$ and the other half to a colder temperature $T_2$. In our case, $T_1$ was 350 K, $T_2$ was 250 K and, thus, we obtained the thermal conductivity around 300 K. Details on the calculation of the thermal conductivity from the time evolution of the temperature difference between the hot and cold regions including convergence considerations are provided in Supplementary Sections S9, S1.4 and S1.5. Again, thermal conductivities are corrected for finite-size effects by calculating differently sized supercells as shown in Supplementary Section S10.

## 4.6 Boltzmann transport equation from anharmonic lattice dynamics (ALD-BTE)

ALD-BTE calculations were performed with the DFT-relaxed as well as with the 300 K unit cells. Phonon band structures were calculated with phonopy[11] with a 2 × 3 × 6 and 3 × 2 × 4 supercell for PE and PT, respectively (see ref. [10] for convergence tests). ALD-BTE calculations were performed with the phonop3y package,[11] with the same settings as in our previous publication, which also provided convergence tests for PE.[10] The default displacement amplitude of 0.03 Å was used in phono3py, for which convergence tests can be found in Supplementary Section S1.7 and ref. [10]. The supercells for calculating the third-order force constants had dimensions of 2 × 2 × 3 and 2 × 2 × 2 for PE and PT, respectively (see ref. [10] for convergence tests). Brillouin zone integration in the ALD-BTE calculation was performed with the tetrahedron method.[84] The meshes for sampling the Brillouin zone were generally chosen, such that the thermal conductivity in chain direction is converged to within 5% (for details see Supplementary Section S1.6). Brillouin zones of PE and PT were sampled with a 4 × 6 × 160 mesh and a 4 × 6 × 48 mesh, respectively.

## 4.7 Boltzmann transport equation with lifetimes from molecular dynamics (MD-BTE)

Simulations to obtain the phonon lifetimes from molecular dynamics (MD-BTE) were performed with the DynaPhoPy package version 1.17.15.[56] We improved the memory efficiency of the code, which allowed the computation of structures with a larger number of atoms in the primitive unit cell (see "Code availability" Section). The required MD simulations were performed in LAMMPS with the MTP$^{MD}$ force fields (see Section 4.2). The used supercells were 2 × 3 × 48 and 2 × 3 × 160 for PT and PE, respectively (see Supplementary Section S1.2 for convergence tests). The MD runs at 300 K were performed for 1 ns (PE) and 2 ns (PT) (following the convergence tests illustrated in Supplementary Figure S1) with a timestep of 0.5 fs. For other temperatures, somewhat larger simulation times were necessary, as detailed in Supplementary Section S3. The velocities were written out for every 10$^{th}$ time step, which had only a negligible impact on the results compared to evaluating the velocities every time step. The resolution of the power spectrum was set to 0.004 THz for PE and 0.002 THz for PT (see Supplementary Section S1.1 for a convergence test). The lifetimes were computed for all symmetry inequivalent grid points to be consistent with the phono3py implementation, which was then employed to obtain the thermal conductivities.

## Data availability

Datasets and LAMMPS scripts generated during this study will be made available on the TU Graz Repository upon publication in a journal.

## Code availability

The modified version of Dynaphopy, that was used in this study, is available at www.github.com/sandrowieser/DynaPhoPy/tree/orig-ez-paper. VASP can be acquired from the VASP Software GmbH (see www.vasp.at); LAMMPS is available at www.lammps.org; MLIP is available at www.mlip.skol-tech.ru/download; the lammps-mlip interface (version 2) is available at www.gitlab.com/ashapeev/interface-lammps-mlip-2; Phonopy is available at www.phonopy.github.io/phonopy; Phono3py is available at www.phonopy.github.io/phono3py.

## Acknowledgements

This research was funded in whole, or in part, by the Austrian Science Fund (FWF) [primarily Grant-DOI: 10.55776/P33903 and in part also 10.55776/P36129]. For the purpose of open access, the authors have applied a CC-BY public copyright license to any author accepted manuscript version arising from this submission. We also acknowledge the Graz University of Technology for support through the Lead Project Porous Materials @ Work for Sustainability (LP-03). Computational results have been obtained using the Vienna Scientific Cluster, VSC-4 and VSC-5.

## Author contributions

Conceptualization, E.Z. and L.R.; methodology, L.R. and S.W.; software, L.R. and S.W.; validation, L.R.; formal analysis, L.R.; investigation, L.R.; resources, E.Z.; data curation, L.R.; writing—original draft preparation, L.R.; writing—review and editing, E.Z., L.R., L.L., and S.W.; visualization, L.R.; supervision, E.Z.; project administration, E.Z.; funding acquisition, E.Z. All authors have read and agreed to the published version of the manuscript.

## Competing interests

The authors declare no competing interests.

**SUPPLEMENTARY INFORMATION**

# Analyzing Heat Transport in Crystalline Polymers in Real and Reciprocal Space

Lukas Reicht[1], Lukas Legenstein[1], Sandro Wieser[1,2], and Egbert Zojer[1,*]

[1] Institute of Solid State Physics, Graz University of Technology, NAWI Graz, Graz, Austria
[2] Institute of Materials Chemistry, TU Wien, Vienna, Austria
* Correspondence: egbert.zojer@tugraz.at

## Table of Contents

# S1 Convergence tests

## S1.1 Simulation time and resolution in MD-BTE calculations

The Boltzmann transport equation calculations with lifetimes from molecular dynamics (MD-BTE) are tested regarding the convergence of the simulation time in the MD simulations used for determining phonon lifetimes. The results are shown in Supplementary Figure S1. As a first step, we here converge the thermal conductivity value at 300 K. The convergence of the mode conductivities of individual phonon bands and for different temperatures is given further below. The calculations are performed with a 2 × 3 × 24 supercell for PT and with a 2 × 3 × 80 supercell for PE. Due to the particularly large supercell for PE, we performed the convergence test for this material with a level 18 MTP, which gives similar results to the level 22 MTP used in the main text (see Supplementary Figure S7). MD-BTE calculations for PT are converged to within 5% after 1.5 ns when using a resolution of 0.004 THz for the power spectrum (see Supplementary Figure S18 for an example of the power spectrum) and for PE an equivalent convergence is reached after 1 ns. Alternatively, at a resolution of 0.002 THz, a simulation time of 1.5 ns is needed to converge the results to within 5%. Regarding the resolution, the difference between 0.002 THz and 0.004 THz is so small, that 0.002 THz can be considered well converged.

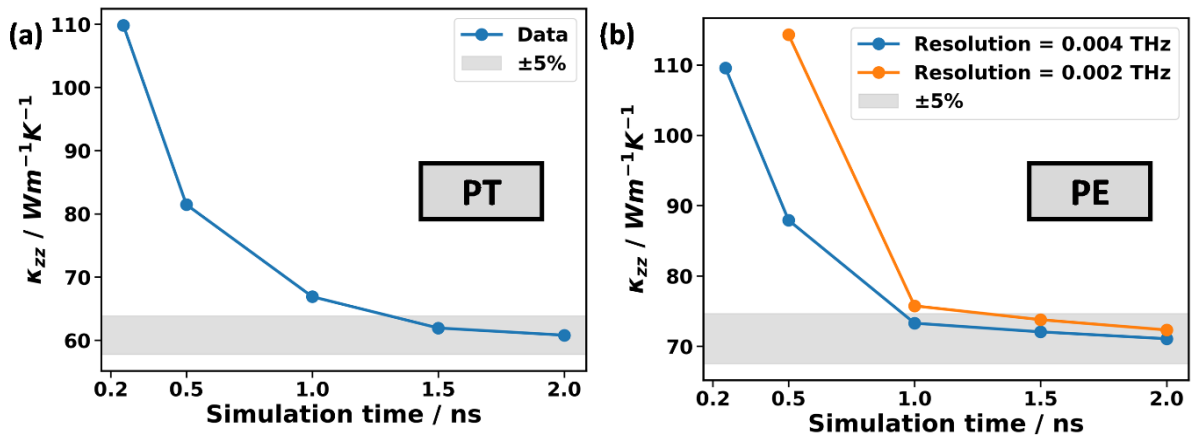

*Figure S1: Thermal conductivity along the chain calculated with MD-BTE with different simulation times for (a) PT and (b) PE. For PT, the resolution of the power spectrum is set to 0.004 THz. For PE, resolutions of 0.002 THz and 0.004 THz are used. For the calculations with the lowest resolution and largest simulation time (i.e., the most converged calculation), a shaded grey region is drawn at ± 5%.*

While the thermal conductivity value at 300 K is sufficiently converged at these simulation times, individual phonon lifetimes, which do not contribute significantly to the thermal conductivity, take a bit longer to converge to a satisfactory degree. To be specific, for PT the mode thermal conductivities are shown for MD-BTE calculations with simulation times of 1 ns, 1.5 ns and 2 ns superimposed on the phonon band structures in Supplementary Figure S2. The 1 ns calculation was performed with a resolution of 0.004 THz, while the 1.5 ns and 2 ns calculations are performed with a 0.002 THz resolution. A 2 × 3 × 48 supercell is used. Evidently, there are qualitative differences for the longitudinal acoustic (LA) mode close to the Γ-point between the 1 ns and the 1.5 ns calculation. It is sensible that differences occur for this mode, since this is the mode with the largest mode thermal conductivity. The 1.5 ns and 2 ns calculations are qualitatively the same, therefore we regard 1.5 ns as sufficient for achieving convergence. In passing we note that the contribution of the "problematic" data points to the thermal conductivity is very small, since the volume in reciprocal space is so small close to the Γ-point (as argued in the main paper). To confirm this, the cumulative thermal conductivity is shown in Supplementary Figure S3.

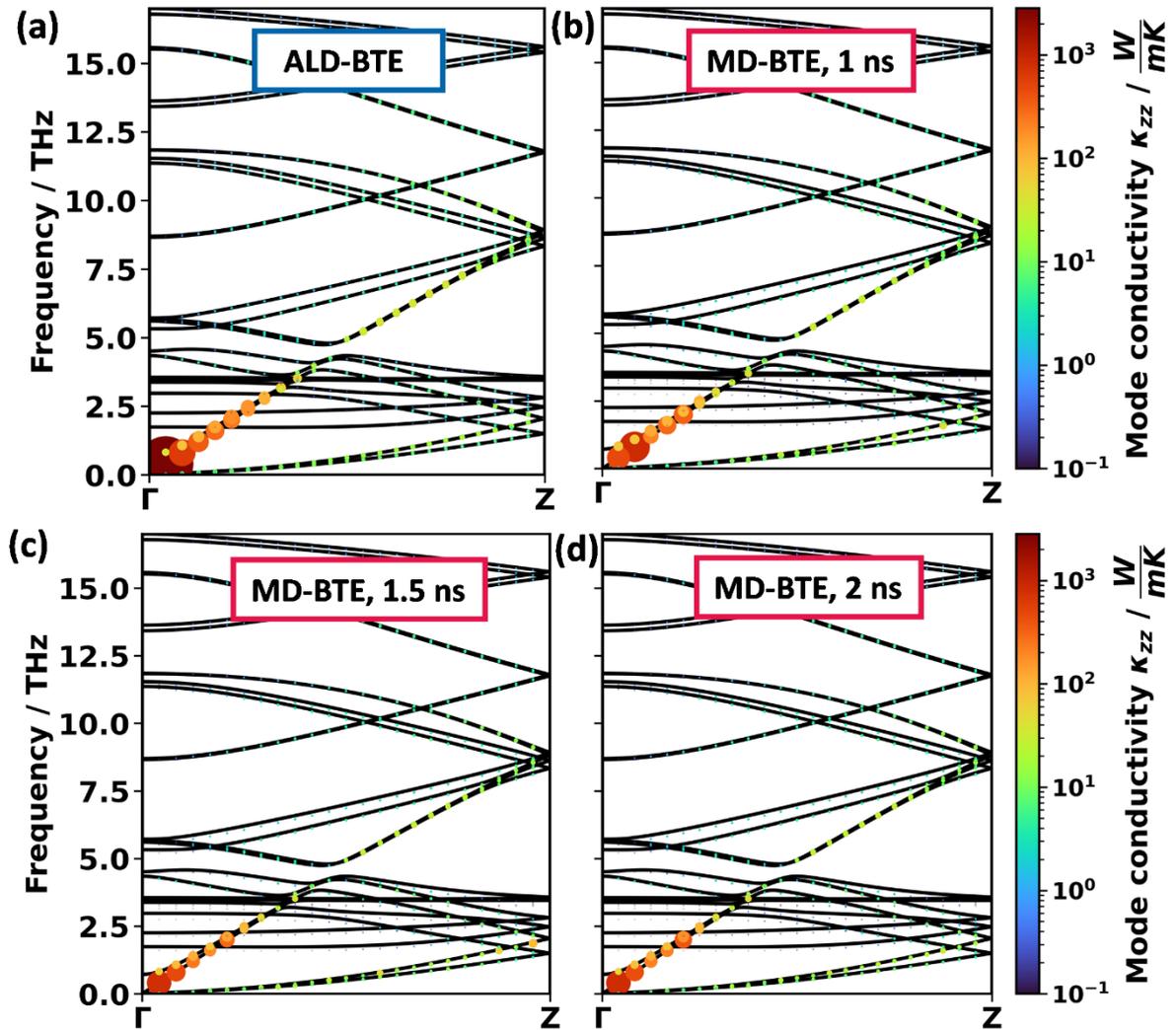

*Figure S2: Phonon band structure of PT, whereby the areas of the data points are linearly enlarged depending on their mode thermal conductivity and colored on a logarithmic scale. In panel (a), the ALD-BTE calculation is shown. The MD-BTE calculation is performed for different simulation times in the MD runs to determine phonon lifetimes in panels (b)-(d).*

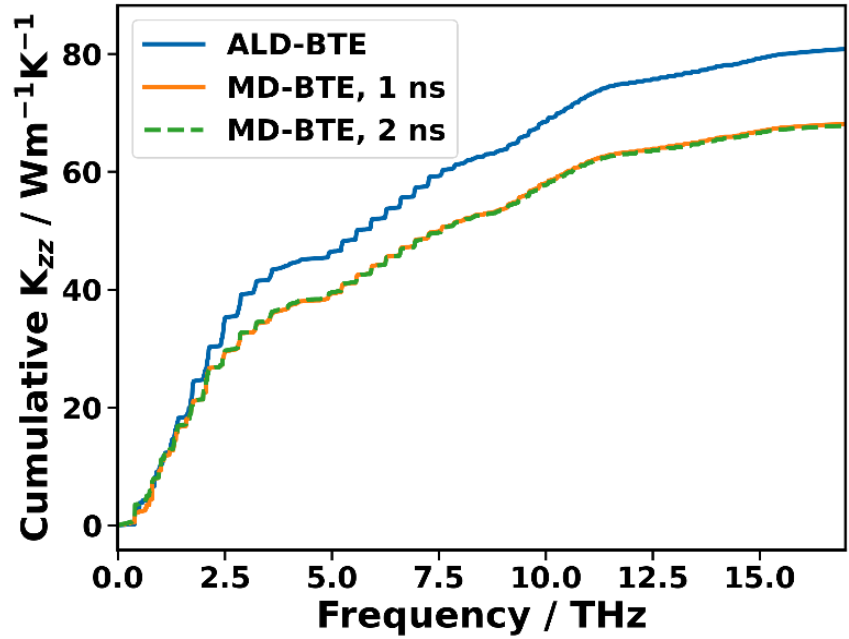

*Figure S3: Cumulative thermal conductivity of PT along the chain direction. For the MD-BTE calculations, simulation times of 1 ns and 2 ns are used, which has only a negligible influence on the cumulative thermal conductivity. The "steps" are due to the finite **q**-mesh sampling of the first Brillouin zone.*

While the MD-BTE simulation of PE at 300 K is very well converged with a simulation time of 1 ns and a resolution of 0.004 THz, this does not guarantee convergence also at other temperatures. Thus, we checked the time convergence for the MD-BTE calculation of PE separately for each temperature. For example, Supplementary Figure S4 shows a convergence test for the MD-BTE calculation at 100 K. As shown in the main paper, the thermal conductivity obtained with MD-BTE is larger at 100 K. This is caused by larger phonon lifetimes, which require longer simulation times and sharper resolutions of the power spectrum. Thus, for calculation at 100 K, 8 ns are necessary when using a resolution of 0.001 THz. At 200 K, already a simulation time of 1 ns and a resolution of 0.004 THz give reasonably well converged results, as shown in Supplementary Figure S5. Still, since the calculation with the longer time was already performed, we used it in the main paper. Therefore, for the results in the main paper, the used simulation times were 8 ns at 100 K, 4 ns at 200 K, and 2 ns at 300 K, 400 K, and 500 K. The resolution was set to 0.001 THz at 100 K, and 0.004 THz at the other temperatures.

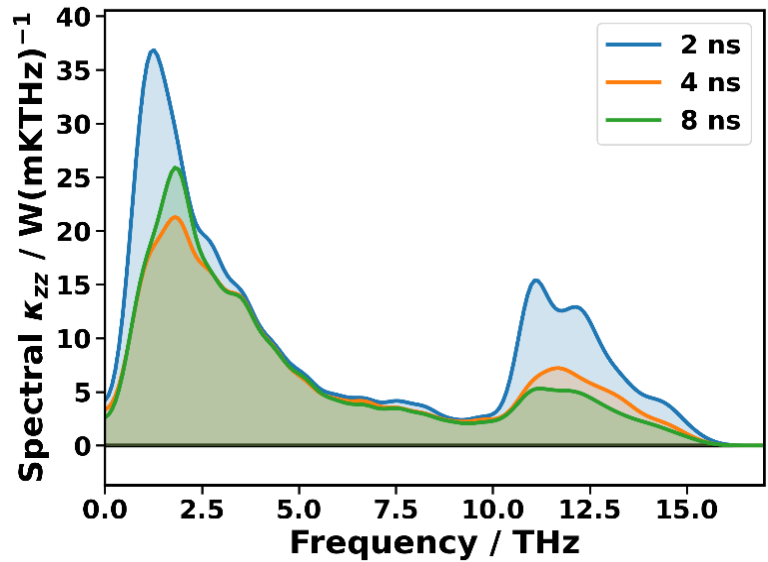

*Figure S4: Spectrally resolved contributions to thermal conductivity of PE at 100 K calculated with MD-BTE for simulation times of 2 ns, 4 ns and 8 ns in the MD runs to calculate phonon lifetimes. The resolution of the power spectrum is 0.001 THz.*

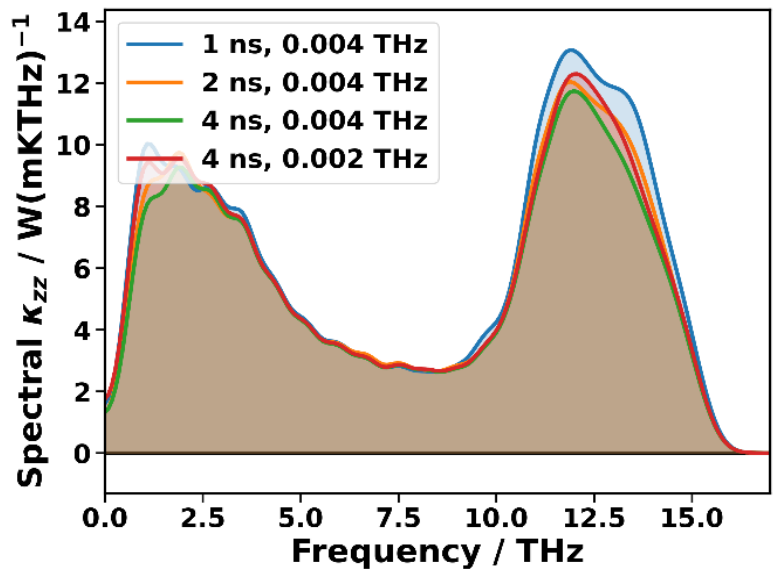

*Figure S5: Spectrally resolved contributions to thermal conductivity of PE at 200 K calculated with MD-BTE. The simulation times are 1 ns, 2 ns and 4 ns with resolutions of 0.004 THz and 0.002 THz as indicated in the figure legend.*

## S1.2 Supercell size in MD-BTE calculations

As a convergence test of the supercell sizes in MD-BTE, the MD-based calculations of phonon lifetimes were performed with different supercells. The resulting cumulative thermal conductivities are shown in Supplementary Figure S6. For all four shown calculations, the simulation time in the MD runs was set to 1 ns and the resolution of the power spectrum was set to 0.004 THz. As in the previous section, the calculations of PE are performed with the level 18 MTP to save computational resources. The converged **q**-meshes from the ALD-BTE calculation (see Supplementary Section S1.6) gave us an initial estimate for the supercell size, which we then tested by doubling it. For PT, the thermal conductivities are 66.9 Wm$^{-1}$K$^{-1}$ and 70.5 Wm$^{-1}$K$^{-1}$ with the 2 × 3 × 24 and 2 × 3 × 48 supercells, respectively. This amounts to a 5% difference. Considering this small difference and that the shapes of the cumulative thermal conductivities are similar, already the 2 × 3 × 24 supercell can be considered converged for calculating the thermal conductivity in chain direction. Since the calculation with the larger 2 × 3 × 48 supercell had already been performed for the present test, this even better converged supercell was used in the discussion in the main paper. Since the unit cell of PT is three times longer than the one of PE, more cell repetitions are required for PE to reach the same supercell length. Correspondingly, supercells of 2 × 3 × 80 and 2 × 3 × 160 were used, which have a similar length than the supercells of PT. For PE, the thermal conductivities are 73.3 Wm$^{-1}$K$^{-1}$ and 77.9 Wm$^{-1}$K$^{-1}$ with supercells of 2 × 3 × 80 and 2 × 3 × 160, respectively. Again, the cumulative thermal conductivities have a similar shape with both supercell lengths. The largest deviations are for phonons around 14 THz to 15 THz. Since the slopes of the cumulative thermal conductivities are similar and the thermal conductivities differ by only 6%, we again regard both supercell sizes as converged in this direction. Moreover, to test the supercell size in the directions perpendicular to the polymer chain, calculations with a supercell size of 2 × 3 × 80 and 4 × 6 × 80 were performed for PE. These show negligible deviations in the phonon lifetimes, spectral thermal conductivity and the thermal conductivity along the chain direction. Thus, we regard the 2 and 3 unit cell repetitions in the **x**- and **y**-directions as converged for calculations of the thermal conductivity along the chain direction.

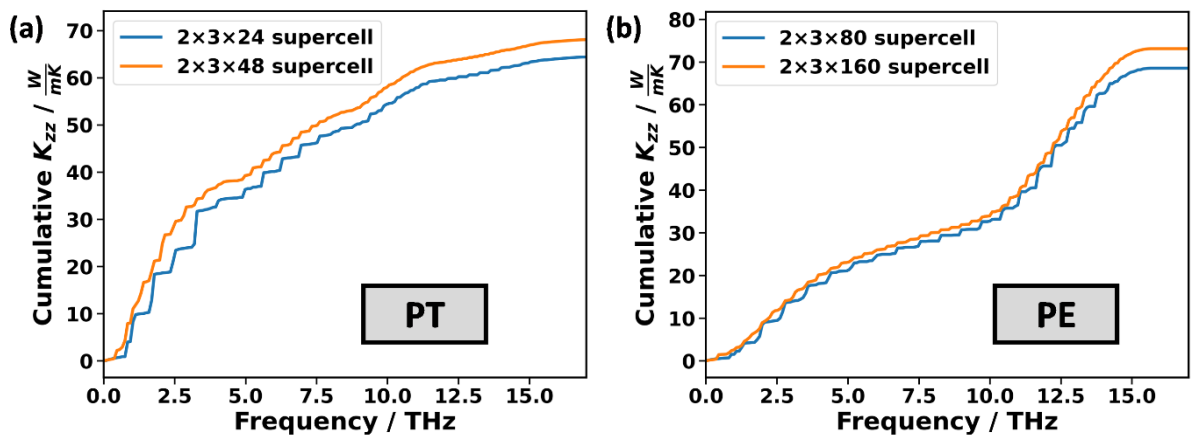

*Figure S6: Cumulative thermal conductivity for different supercell sizes of (a) PT and (b) PE.*

## S1.3 Level of the MTP

In our previous publication[1] and in the Methods section of the main paper, we suggested a protocol for parametrizing an MTP meant for molecular dynamics simulations and referred to it as MTP$^{MD}$. It is trained on data sampled at 15 K to 500 K and has a level of 22. This MTP$^{MD}$ is used for the MD-BTE calculation. To test the influence of the level, we parametrized five MTPs for PE with level 18 on the same training data that were used for the level 22 MTP$^{MD}$. The cumulative thermal conductivities calculated with the respective "best" MTPs are shown in Supplementary Figure S7. Qualitatively, they agree very well. Also quantitatively the agreement is highly satisfactory: The corresponding thermal conductivities are 79 Wm$^{-1}$K$^{-1}$ and 84 Wm$^{-1}$K$^{-1}$ with the level 18 and level 22 MTPs, respectively. This difference is similar to the difference between the five individual MTP$^{MD}$s (see Supplementary Section S18). In light of this very good agreement, we conclude that already the 18 MTP would be sufficiently accurate. In view of the results by Wu et al., who recently suggested that inaccuracies of machine learned potentials can lead to an underestimation of the thermal conductivity,[2] we picked the level 22 MTP for all calculations in the main manuscript. In this way we are confident that the accuracy of the used MTP is so high that there is no need for applying the correction also suggested by Wu et al.[2].

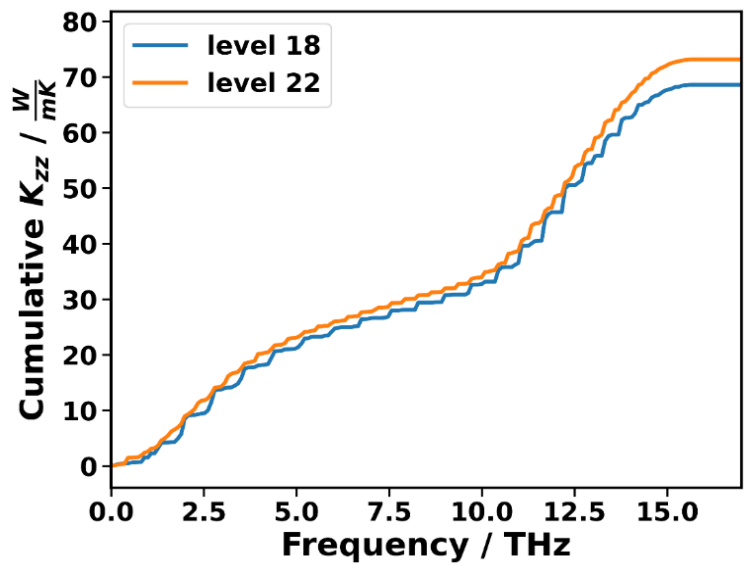

*Figure S7: Cumulative thermal conductivity of PE calculated with MD-BTE using MTPs with level 18 and level 22. The supercell size is 2 × 3 × 160 and the simulation time when determining phonon lifetimes is 1 ns.*

## S1.4 Simulation time in AEMD

In AEMD simulations, the temperature difference decays as shown in Supplementary Figure S8a for an example calculation on PT. The thermal conductivity is evaluated for different simulation times and its convergence is plotted in Supplementary Figure S8b. The thermal conductivity peaks at around 75 ps at a value differing by less than 5% from the value at 200 ps. Therefore, the thermal conductivity for times beyond 75 ps can be regarded as converged for this calculation, yet the even more converged 200 ps were used for the data reported in the main manuscript. For the calculation in Supplementary Figure S8b, the unit cell was 256 repetitions long along the chain direction. When performing calculations with larger unit cells, the simulation time was scaled linearly with the length of the unit cell. This resulted in calculations that show a similar degree of convergence. Accordingly, the used calculation times are 100 ps, 150 ps, 200 ps, 300 ps and 400 ps for unit cell repetition of 128, 192, 256, 384 and 512 times. The same analysis as just described for PT was also performed for PE. For PE, the simulation times are 60 ps, 100 ps, 150 ps, 200 ps and 300 ps for unit cell repetition of 432, 720, 1080, 1440 and 2160 times. As mentioned above, more unit cell repetitions are needed for PE, since its unit cell is three times shorter than the one of PT.

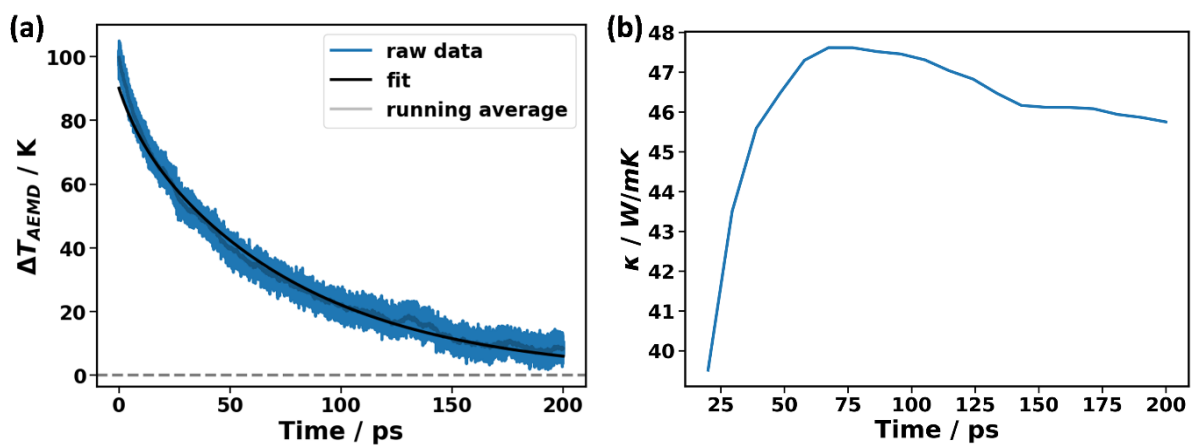

*Figure S8: Panel (a): Temperature difference between hot and cold regions, $\Delta T_{AEMD}$, decaying with time in an AEMD simulation of PT. The raw data is shown in blue and a running average in light grey. The fit that is used to evaluate the thermal conductivity is shown as a black line. Panel (b): Simulation time convergence of the thermal conductivity in an AEMD simulation of PT.*

### S1.5 Number of exponentials in fit of AEMD data

The temperature difference in AEMD simulations is fitted by a sum of exponentials, as detailed in Supplementary Section S9. The thermal conductivity needs to be converged with respect to the number of exponentials, denoted as $N_{exp}$. As shown in Supplementary Figure S9, three exponentials are enough to reach a very convincing convergence. Therefore, three exponentials were used throughout the manuscript.

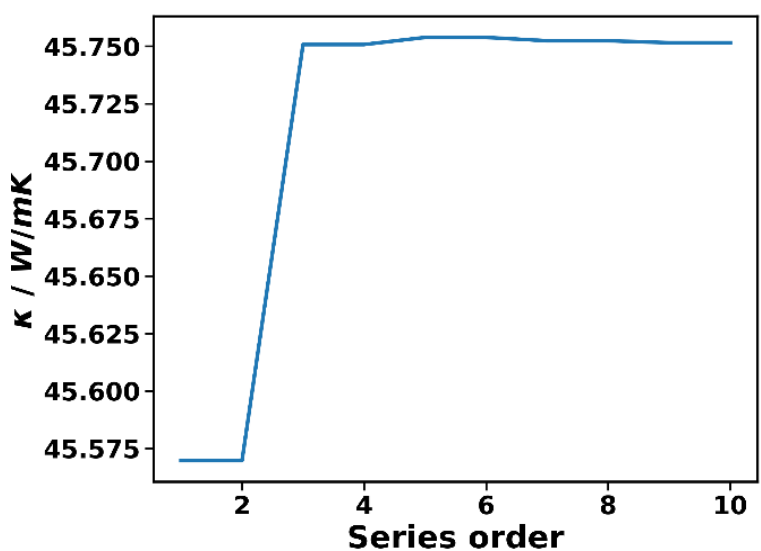

*Figure S9: Convergence of the thermal conductivity with respect to the number of exponentials $N_{exp}$ in the fit function.*

## S1.6 Q-mesh in ALD-BTE calculations

In ALD-BTE calculations, the Brillouin zone is sampled with a **q**-mesh, which has to be checked for convergence. For PT, a convergence test for the **q**-points along the chain direction is shown in Supplementary Figure S10a. The directions perpendicular to the chain were investigated in a separate convergence test. The thermal conductivity along the chain direction $\kappa_{zz}$ (with the full BTE and in RTA) is converged to within 5% for a 4 × 6 × 26 **q**-mesh. We note that this calculation was performed with an MTP that was parametrized in a slightly different way than the MTPs of the main manuscript. Namely, this MTP has a level of 28 and was trained on 169 training structures that were sampled at 15 K to 300 K. Since this MTP yields a similar thermal conductivity as the MTPs used in the main paper, we regard the results obtained for the **q**-mesh convergence test to be transferable also to the other MTPs. Therefore, the **q**-mesh convergence test was not repeated for the MTPs of the main paper.

For PE with the DFT-relaxed unit cell, a convergence test of the **q**-mesh can be found in the Supplementary Materials of Ref. [1]. The convergence behavior for the 300 K unit cell is shown in Supplementary Figure S10b. There, the **q**-points in **x**- and **y**-direction are held fixed at 4 and 6 points, while the number of **q**-points along the chain direction is varied. The numerical values corresponding to the data in Supplementary Figure S10b are listed in Table S1. For the RTA, the convergence is rather benign and already the thermal conductivity with a **q**-mesh of 4 × 6 × 80 differs by less than 5% from the values obtained for larger **q**-meshes. For the full BTE simulation, the situation is somewhat more involved with an outlier for 200 **q**-points. We opted to use the 4 × 6 × 160 **q**-mesh for all calculations of PE, since it is reasonably well converged with a roughly 5% difference in the thermal conductivity as compared to the 4 × 6 × 320 **q**-mesh, while remaining computationally manageable.

Alongside these values, we also list the total number of eigenvalues in the collision matrix and the number of negative eigenvalues in the collision matrix. A definition of the collision matrix can be found in equation (61) of Ref. [3]. From that equation it is evident that the collision matrix is positive semidefinite, as long as all phonon frequencies are positive. We checked explicitly that in fact all phonon frequencies are positive. Therefore, we conclude that any negative eigenvalues of the collision matrix are a numerical artifact. The absolute values of these negative eigenvalues are comparable or smaller than the arithmetic mean of the positive eigenvalues. Since there is only a very small number of negative eigenvalues compared to positive eigenvalues, we argue that these negative eigenvalues have no meaningful impact on the calculated thermal conductivities. We observe negative eigenvalues for roughly half of our simulations, including the simulations of PT. Their appearance seems random. For example, when parametrizing five MTPs for PT with different initialization, around half of them produce negative eigenvalues, while the corresponding five thermal conductivities are all in good agreement with each other. The appearance of negative eigenvalues does not at all correlate with whether the respective thermal conductivity values are reasonable. In particular, it does not explain the unreasonable thermal conductivities that are observed for the 300 K unit cell with PE when MTPs are used (see Supplementary Section S18 for more details).

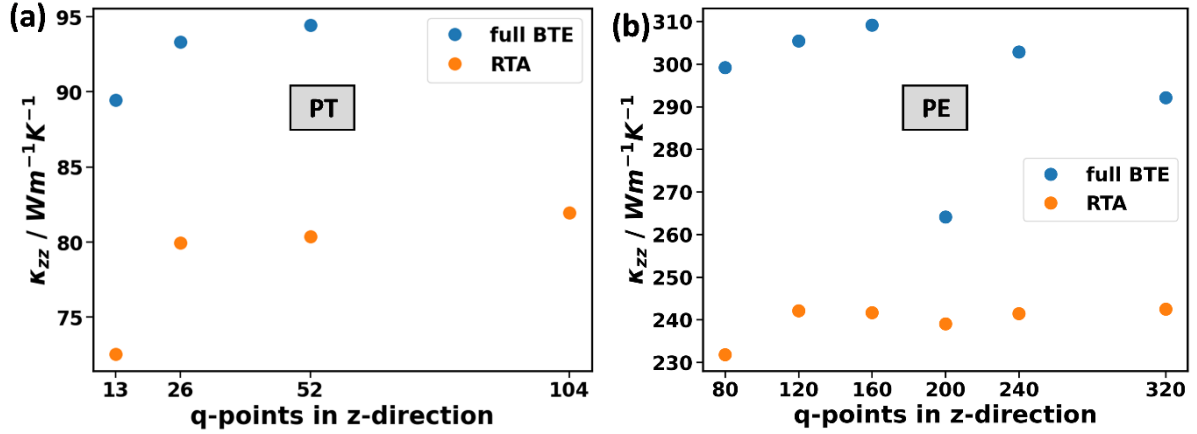

*Figure S10:* **Q**-mesh convergence test along the chain direction for (a) PT and (b) PE. The calculation for PT was performed with the DFT-relaxed unit cell using an MTP, which was parametrized in a slightly different way than the MTPs of the main manuscript (see main text for details). For PE, the DFT calculation for the 300 K unit cell is shown.

*Table S1: Numerical values for the convergence test depicted in Supplementary Figure S10b. Additionally, we list the total number of eigenvalues in the collision matrix and the number of negative eigenvalues in the collision matrix.*

| q-mesh | 4 × 6 × 80 | 4 × 6 × 120 | 4 × 6 × 160 | 4 × 6 × 200 | 4 × 6 × 240 | 4 × 6 × 320 |
| --- | --- | --- | --- | --- | --- | --- |
| $\kappa_{zz}$ in RTA / $Wm^{-1}K^{-1}$ | 231.8 | 242.0 | 241.6 | 239.0 | 241.4 | 242.4 |
| $\kappa_{zz}$ full BTE / $Wm^{-1}K^{-1}$ | 299.1 | 305.4 | 309.1 | 264.1 | 302.8 | 292.1 |
| Total number of eigenvalues | 53,136 | 79,056 | 104,976 | 130,896 | 156,816 | 208,656 |
| Number of negative eigenvalues | 3 | 3 | 4 | 4 | 4 | 5 |

### S1.7 Displacement amplitude in ALD-BTE calculations

In phono3py calculations, atoms are by default displaced by 0.03 Å for calculating third-order force constants. This displacement amplitude can be controlled with the "amplitude" tag in phono3py. For too small displacement amplitudes, the numerical noise becomes too large, while for too large amplitudes, higher order anharmonicities play an increasing role and "distort" the calculated third order force constants. Thus, the goal is to find a region, in which the magnitude of the displacement does not severely impact the results. To test the convergence of the displacement amplitude, values of 0.01 Å, 0.03 Å, 0.05 Å, 0.07 Å, 0.09 Å and 0.11 Å were used. Examples of such convergence tests can be seen in Supplementary Figure S11 for the DFT-relaxed cell of PE and the 300 K cell of PT. These tests were performed for the DFT-relaxed as well as the 300 K unit cells. They were also performed in the vdW-bonded directions, respectively for the Peierls contribution, phonon tunneling contribution and total thermal conductivity (see Supplementary Section S3 for details on the Peierls and phonon tunneling contribution). Since showing all of these plots would be rather lengthy, we instead describe the outcome here in the text and show two representative plots in Supplementary Figure S11. In all of these scenarios, the thermal conductivities for the displacement amplitude of 0.03 Å and 0.05 Å agreed well. On the one hand, the displacement amplitude of 0.01 Å sometimes seemed too small for the vdW-bonded directions and also in Supplementary Figure S11b, as it deviates from the larger amplitudes. On the other hand, displacement distances beyond 0.05 Å tend to give thermal conductivities that deviate from the values at 0.03 Å and 0.05 Å. Thus, considering all the performed convergence tests, the displacement amplitude of 0.03 Å was found to give the best convergence. This displacement amplitude was also used in the calculations of the main manuscript. In addition to these comprehensive tests with the MTPs, the DFT calculation of PE for the DFT unit cell was performed with an amplitude of 0.03 Å and 0.05 Å, yielding very similar results (see ref. [1]).

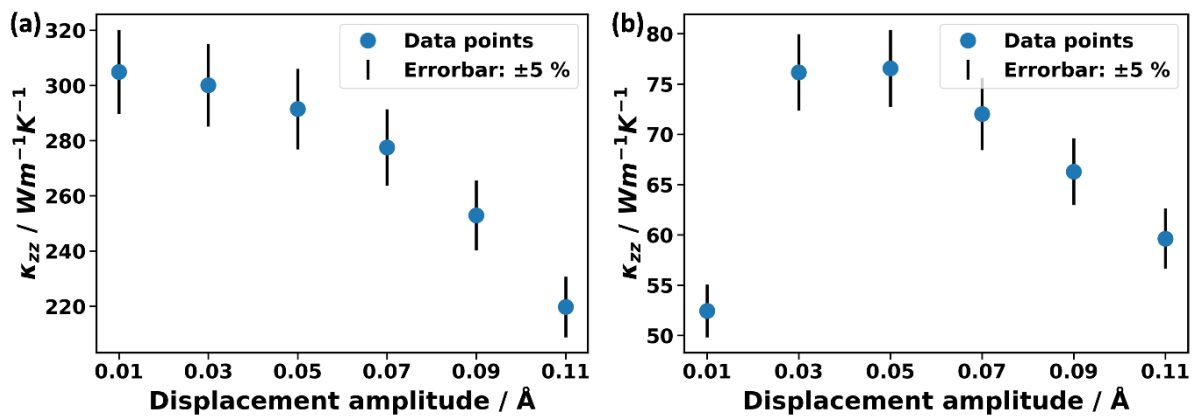

Figure S11: Convergence of $\kappa_{zz}$ with respect to the displacement amplitude in phono3py. Panel (a) depicts the MTP-calculated total thermal conductivity for the DFT-relaxed cell of PE. Panel (b) shows the calculation of PT with the 300 K cell. Since we applied a convergence criterion of 5% to the **q**-mesh, error bars of that magnitude are drawn to put the deviations into perspective.

## S2 Comparison with existing literature

### S2.1 Comparison of polythiophene's thermal conductivity to the simulation results of Cheng et al.

Cheng et al. calculated the thermal conductivity of crystalline PT using the temperature dependent effective potential (TDEP) method and obtained a thermal conductivity of 198 Wm$^{-1}$K$^{-1}$.[4] That result is rather different from the thermal conductivities reported in the main paper of the present study. To address this issue, we identified four possible reasons for this difference: These are the structure (herein referred to as "Cheng's structure" and "our structure"), the vdW-correction (Cheng used the original vdW-DF[5], while we used PBE+D3[6–8]), the different method (Cheng used the TDEP method, while we used ALD-BTE, MD-BTE, AEMD and NEMD) and various convergence settings. Upon request, the authors of the paper were so kind as to send us the structure that they used and details on their calculation. The structures differ insofar as in our structure one of the PT chains is shifted along the chain axis, thereby reducing the symmetry. This intentional breaking of the symmetry is crucial such that the structure does not get stuck on a saddle point of the potential energy surface. Indeed, when we calculate phonons with Cheng's structure and the PBE+D3 functional, we find a negative Γ-mode that corresponds to a translation of one of the PT chains in the unit cell relative to the other. This is precisely the chain shift that entails the difference between our structure and Cheng's structure. A displacement along such a negative Γ-mode corresponds to a lowering in energy, thereby showing that Cheng's structure is not in an energy minimum (at least for the PBE+D3 functional). To investigate the matter further, we relax both structures with each of the DFT functionals. The resulting energy and volume can be found in Table S2 and Table S3. For both functionals, the structure with shifted polymer chains (i.e., "our structure") has a lower energy. These results suggest that "our structure" should be preferred, especially for calculations of phonons and BTE calculations. Notably, the energy barrier between our structure and Cheng's structure is very low as can be inferred from the observation that, when performing molecular dynamics simulations in VASP, a shift of the chains relative to each other is observed. This creates a certain level of disorder with respect to the shift in chain direction. At this point, we can only speculate to what extent the differences in the structures are relevant when applying the TDEP method, as for room temperature thermal displacements such shifts are conceivable.

*Table S2: Total energy in meV/atom for the DFT functionals PBE+D3 [6–8] and vdW-DF [5], calculated for optimizations of the "Cheng's structure" and "our structure" (for details see main text).*

|                    | PBE+D3                        | vdW-DF                        |
|--------------------|-------------------------------|-------------------------------|
| **Cheng's structure** | 3.7 meV/atom                | 2.2 meV/atom                  |
| **Our structure**  | 0 meV/atom (by normalisation) | 0 meV/atom (by normalisation) |

*Table S3: The volume for the same calculations as in Table S2. For comparison, the experimentally determined volume from Mo et al. [9] is given.*

|                    | PBE+D3  | vdW-DF  |
|--------------------|---------|---------|
| **Cheng's structure** | 332 Å$^3$ | 362 Å$^3$ |
| **Our structure**  | 324 Å$^3$ | 367 Å$^3$ |
| **EXPERIMENT** [9] | 347 Å$^3$          ||

In addition to comparing energies in Table S2, we compare volumes in Table S3. The non-negligible difference in cell volumes is primarily caused by the choice of the DFT functional, while the choice of the starting structure for the relaxation ("our structure" vs. "Cheng's structure") has only a minor impact. This raises the question as to which of the DFT functionals produces unit cells that better fit experiments. To make this comparison, lattice constants are compared in Table S4 with experiments. Comparing the DFT results with experiments by Mo et al. and Brückner et al.[9,10] for each of the three lattice constants, it can be seen that for $a_1$, PBE+D3 gives a superior agreement. For $a_2$, vdW-DF gives better agreement, with the experimental value lying between the DFT results. For $a_3$, vdW-DF also agrees better to experiment. Here, it is important to consider that the DFT-calculated values correspond to 0 K, while the experiments (at least the one by Brückner et al.[10]) was performed at room temperature. PT exhibits thermal expansion in **x**- and **y**-direction (see Supplementary Section S12). Thus, it is more reasonable that DFT underpredicts the experimental lattice constants, as is the case for the PBE+D3 functional. Overall, the agreement to experiment is good with both functionals and starting structures. In passing we note that the "best" MTP gives very similar values to DFT, more similar than the difference caused by the functional.

*Table S4: Lattice constants $a_1$, $a_2$, and $a_3$ are calculated with different DFT functionals (PBE+D3 [6–8], vdW-DF [5]) and starting structures, and are compared to experiments by Mo et al. and Brückner et al.[9,10] The measurement of Brückner et al. [10] was performed at room temperature, while for the experiment of Mo et al. [9] we could not find a specified temperature. The lattice constants $a_1$, $a_2$, and $a_3$ correspond to the unit cell lengths in **x**-, **y**- and **z**-direction as shown in Figure 1 of the main paper. The lattice constants with "our structure" and PBE+D3 reported here are identical to those contained in our previous publication.[1]*

|  | $a_1$ [Å] | $a_2$ [Å] | $a_3$ [Å] |
| --- | --- | --- | --- |
| **Theory** | | | |
| Our structure, PBE+D3 | 5.542 | 7.530 | 7.785 |
| Cheng's structure, PBE+D3 | 5.639 | 7.561 | 7.786 |
| Our structure, vdW-DF | 5.840 | 7.900 | 7.856 |
| Cheng's structure, vdW-DF | 5.908 | 7.929 | 7.856 |
| **Experiment** | | | |
| Experiment[9] | 5.55 | 7.80 | 8.03 |
| Experiment at room temperature[10] | 5.33 | 7.79 | - |

In summary, the calculated thermal conductivity of Cheng et al. [4] is twice as large as our result which we attribute to one of the following reasons: (1) different functional, (2) structure of Cheng et al. is on a saddle point of the potential energy surface (3) different methods for calculating the thermal conductivity (phono3py vs. TDEP). In absence of experimental values, we cannot definitively assess which calculation should be preferred.

### S2.2 Comparison of polyethylene's thermal conductivity with results of Wang et al.

Wang et al. report a thermal conductivity of 237 Wm$^{-1}$K$^{-1}$ for the PE crystal.[11] Since they used a DFT-relaxed unit cell, similar to the DFT-relaxed unit cell in this work, we compare their result with our result obtained with the DFT-relaxed unit cell in Table S5. There one can see that the thermal conductivity of Wang et al. significantly deviates from our result. There are multiple reasons that could explain this difference: Firstly, different DFT functionals were used. Wang et al. used the local-density approximation (LDA) with the optB88-vdW[12,13] correction, while we use PBE+D3 [6–8]. Secondly, different energy cutoffs were used. Wang et al. used an energy cutoff of 550 eV, while we use 700 eV. Thirdly, we used larger meshes than Wang et al. in the BTE calculation. They show a convergence test in Supplementary Figure S4 of their Supplementary Information, where the thermal conductivity remains unchanged for a mesh of 40 and 50 points. This is in contrast to our convergence tests shown in Ref. [1], where we find that larger meshes are necessary. We found that 60 points are sufficient for the BTE in the RTA, while for the full BTE at least 80 points are necessary. It is conceivable, that Wang's thermal conductivity would be increased with a larger mesh.

Table S5: Thermal conductivity of PE crystal along the chain direction at 300 K as calculated in this study is compared to the results of Wang et al.[11] . Values from this study are calculated with the DFT-relaxed unit cell.

|  | $\kappa_{zz}$ from Wang et al. [11] / Wm$^{-1}$K$^{-1}$ | $\kappa_{zz}$ from this work / Wm$^{-1}$K$^{-1}$ |
|---|---|---|
| **RTA** | ~200 | 296 |
| **full ALD-BTE** | 237 | 398 |

## S3 Phonon tunneling contribution to the thermal conductivity

Simoncelli et al. introduced a unified theory for thermal transport in crystals and glasses.[14] In that theory, in addition to the Peierls thermal conductivity from the Boltzmann transport equation, they proposed to add a second term, labelled coherences' thermal conductivity $\kappa_C$, that accounts for phonon tunnelling.[14] This contribution is added to the Peierls thermal conductivity $\kappa_P$ to yield the total thermal conductivity $\kappa_T$. These three quantities are given in Table S6 through Table S9 for PT and PE with the DFT-relaxed and 300 K cells, respectively. For both materials and cells, the contribution from the coherences' thermal conductivity is negligible in chain direction. Therefore, and since it complicates a mode-by-mode analysis, $\kappa_C$ is neglected in the main paper. In the vdW-bonded directions, however, $\kappa_C$ has a much larger relative contribution. In the most extreme case, which is PE with the 300 K cell, $\kappa_C$ even accounts for 38% and 33% of the total thermal conductivity in the respective vdW-bonded directions.

Table S6: Peierls thermal conductivity, $\kappa_P$, coherences' thermal conductivity, $\kappa_C$, and total thermal conductivity, $\kappa_T$, of PT in Wm$^{-1}$K$^{-1}$. The RTA is employed, and the Brillouin zone is sampled with a 7 × 10 × 26 **q**-mesh. The DFT-relaxed cell is used and the MTP$^{phonon}$ is trained on data sampled at 15 K to 100 K in an NpT ensemble (as described in the Methods section of the main paper). For the definition of the different directions see Figure 1 of the main paper.

|  | x-direction / Wm$^{-1}$K$^{-1}$ | y-direction / Wm$^{-1}$K$^{-1}$ | z-direction / Wm$^{-1}$K$^{-1}$ |
|---|---|---|---|
| $\kappa_P$ | 0.496 | 0.428 | 83.2 |
| $\kappa_C$ | 0.045 | 0.025 | 1.6 |
| $\kappa_T$ | 0.541 | 0.453 | 84.8 |

*Table S7: Same as Table S6, but with the 300 K cell and the respective MTP that is trained on data sampled at 15 K to 100 K with the unit cell fixed to the 300 K unit cell (as described in the Methods section of the main paper).*

|  | x-direction / Wm$^{-1}$K$^{-1}$ | y-direction / Wm$^{-1}$K$^{-1}$ | z-direction / Wm$^{-1}$K$^{-1}$ |
|---|---|---|---|
| $\kappa_P$ | 0.527 | 0.464 | 73.6 |
| $\kappa_C$ | 0.037 | 0.022 | 1.9 |
| $\kappa_T$ | 0.564 | 0.485 | 75.5 |

*Table S8: Peierls thermal conductivity, $\kappa_P$, coherences' thermal conductivity, $\kappa_C$, and total thermal conductivity, $\kappa_T$, of PE in Wm$^{-1}$K$^{-1}$. The RTA is employed, and the Brillouin zone is sampled with a 10 × 15 × 160 **q**-mesh. The calculations are performed using DFT with an energy cutoff of 700 eV and with the DFT-relaxed unit cell.*

|  | x-direction / Wm$^{-1}$K$^{-1}$ | y-direction / Wm$^{-1}$K$^{-1}$ | z-direction / Wm$^{-1}$K$^{-1}$ |
|---|---|---|---|
| $\kappa_P$ | 0.543 | 0.460 | 306.1 |
| $\kappa_C$ | 0.049 | 0.030 | 0.7 |
| $\kappa_T$ | 0.592 | 0.490 | 306.8 |

*Table S9: same as Table S8, but with the 300 K unit cell. DFT with an energy cutoff of 700 eV is used.*

|  | x-direction / Wm$^{-1}$K$^{-1}$ | y-direction / Wm$^{-1}$K$^{-1}$ | z-direction / Wm$^{-1}$K$^{-1}$ |
|---|---|---|---|
| $\kappa_P$ | 0.072 | 0.097 | 230.1 |
| $\kappa_C$ | 0.044 | 0.047 | 0.4 |
| $\kappa_T$ | 0.117 | 0.144 | 230.5 |

Surprisingly, for PT, the 300 K cell yields a larger thermal conductivity in vdW-bonded directions by 4% and 7%. However, it needs to be stressed here that these calculations were performed with different MTPs for the respective unit cells. When calculating the thermal conductivity with five MTPs for the DFT cell they show a spread of 12% and 15% in the vdW-bonded directions. Thus, we attribute the apparent increase in thermal conductivity to the spread of the MTPs. In fact, when calculating the thermal conductivity for the DFT-relaxed and 300 K cell with the same MTP, we see a decrease in thermal conductivity for the 300 K cell. This is consistent with our experience that a larger unit cell leads to a reduced thermal conductivity.

On this topic, it is also worthwhile mentioning the significant decrease of the thermal conductivity perpendicular to the PE chains for the 300 K cell. Qualitatively, this result is consistently obtained also when using MTPs. The much larger change of the perpendicular thermal conductivity in PE than in PT between the two cells correlates with a significantly more pronounced thermal expansion of PE (see Supplementary Section S12). A detailed discussion of the atomistic origin of this thermal conductivity increase goes beyond the scope of the present paper.

## S4 Comparing ALD-BTE and AEMD results for the thermal conductivity in a vdW-bonded direction

While the focus of the main paper is on the thermal conductivity along the chain direction, we also performed AEMD simulations in one of the vdW-bonded directions of PT. In passing we note that this requires a new finite-size extrapolation. AEMD is compared with the ALD-BTE result for the 300 K unit cell in Table S10. The AEMD result of 0.60 Wm$^{-1}$K$^{-1}$ agrees well with the ALD-BTE result of 0.56 Wm$^{-1}$K$^{-1}$. Notably, including the coherences' thermal conductivity clearly improves the agreement between the two calculations.

Table S10: Thermal conductivity of PT in the vdW-bonded **x**-direction calculated with ALD-BTE and AEMD. The ALD-BTE calculation is the same as the one reported in Table S7, with settings described there. Uncertainties in AEMD are the standard deviations (68% confidence interval) from the finite-size extrapolation.

|  | $\kappa_{xx}$ / Wm$^{-1}$K$^{-1}$ |
| --- | --- |
| **ALD-BTE, $\kappa_P$** | 0.527 |
| **ALD-BTE, $\kappa_C$** | 0.037 |
| **ALD-BTE, $\kappa_T$** | 0.564 |
| **AEMD** | Zaoui fit: 0.60 ± 0.03 |
|  | Linear fit: 0.68 ± 0.02 |

## S5 Effect of phonon renormalization

Higher order phonon scatterings, especially four-phonon scatterings, lead to a renormalization of phonon frequencies.[15] To study the impact of renormalization, phonon frequencies are calculated with Dynaphopy at commensurate **q**-points. Within Dynaphopy, renormalized force constants are calculated from the phonon frequencies at commensurate **q**-points, which allows to interpolate between the commensurate **q**-points.[16] The renormalization leads to a shift in phonon frequencies as shown in Supplementary Figure S12 for PE. Here, we used molecular dynamics at 300 K, an MTP$^{MD}$ and Dynaphopy to calculate the renormalized band structure with a 2 × 3 × 80 supercell and 250 ps simulation time. The longitudinal acoustic band, which is the main carrier of heat, remains unchanged upon renormalization, while other modes shift somewhat. There are a few deficiencies of the renormalized phonon band structure: Two optical bands have a "bump" at 10 THz, which appears to be an artefact of the calculation, since such bumps seem unphysical in the absence of any avoided crossings. Furthermore, to properly converge the phonon band structure calculated with Dynaphopy, larger supercells than for the phonopy calculations are required. Taking PE as an example, a 2 × 3 × 6 supercell yields a converged phonon band structure with phonopy, while for a Dynaphopy calculation, such a supercell size yields phonon bands that are wildly fluctuating and are clearly not converged. Even for the 2 × 3 × 80 supercell (see Supplementary Figure S12), the transverse acoustic bands along the Γ-Z path have a slight "wave-like" dispersion, which is clearly unphysical. When increasing the supercell size even further to 4 × 6 × 80, this "wave-like" character is greatly reduced. However, the "bump" at 10 THz remains even for the larger supercell. Also, when doubling the simulation time from 250 ps to 500 ps, it does not vanish. In passing we note that these huge supercells drastically increase the memory demand and computational cost for obtaining the phonon band structure.

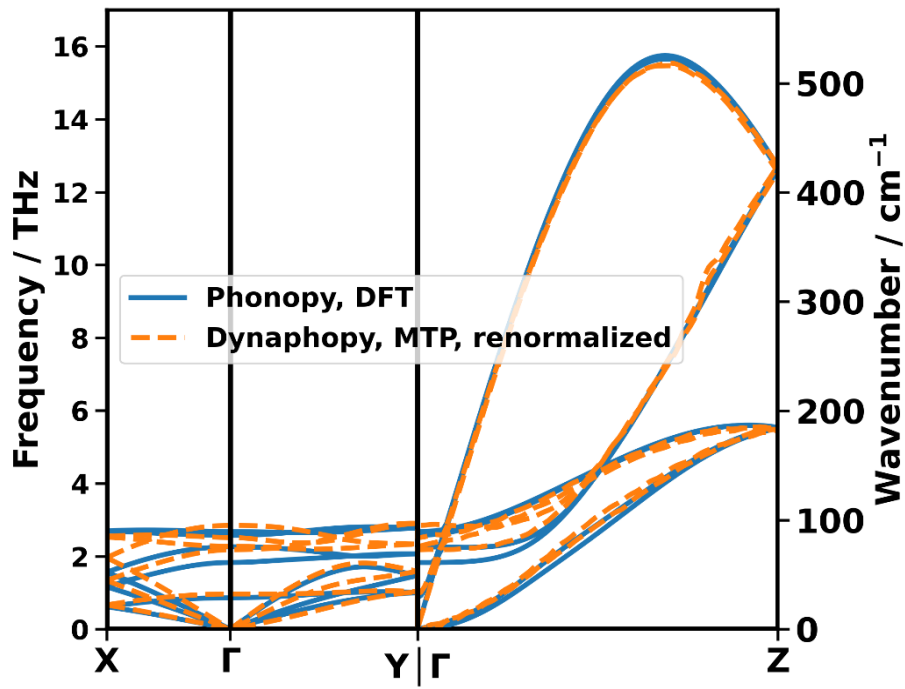

*Figure S12: Phonon band structure of PE. The band structure shown as blue solid lines is calculated with finite displacements using phonopy and DFT. The calculation resulting in the phonon bands shown as dashed orange line was performed using Dynaphopy and an MTP. The Dynaphopy calculation captures phonon renormalization at 300 K.*

To investigate the influence on thermal conductivity, the renormalized second order force constants were used to perform an ALD-BTE calculation with phono3py. The calculation was performed with a 10 × 15 × 160 **q**-mesh in the RTA, the (not renormalized) third order force constants from phono3py and the renormalized second order force constants of the 2 × 3 × 80 supercell. The thermal conductivity $\kappa_{zz}$ is 230 Wm$^{-1}$K$^{-1}$ with the unrenormalized force constants and 217 Wm$^{-1}$K$^{-1}$ with the renormalized force constants. We regard this 6% difference as an indication that phonon frequency renormalization does not play an important role here. This 6% difference appears especially small, when we consider that the phonon band structures in the two cases have been obtained based on fundamentally different approaches, which can also contribute to the observed minor deviation. Also for PT we observe that the renormalization typically lowers the thermal conductivity by around 5%. To make also a qualitative assessment, the spectral $\kappa_{zz}$, which is the derivative of the cumulative thermal conductivity, is shown in Supplementary Figure S13. It shows reasonable agreement of the ALD-BTE calculations with and without renormalization. The renormalized calculation has a somewhat smaller thermal conductivity at around 12-14 THz, while it has a larger thermal conductivity at around 2-5 THz. Since the renormalization effects are small, and considering the issues in the calculations of the renormalized band structures discussed above, phonon renormalization is not considered in any calculations except the ones present in the current section.

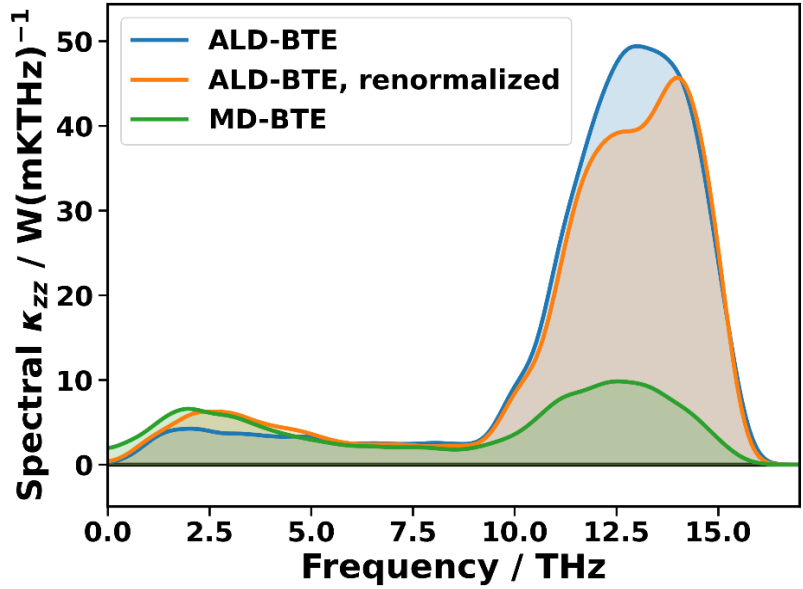

*Figure S13: Spectrally resolved contributions to thermal conductivity of PE along the polymer chain, calculated as the derivative of the cumulative thermal conductivity. The ALD-BTE and MD-BTE calculation are the same as in the main paper. The "ALD-BTE, renormalized" calculation is performed with phono3py, whereby the renormalized second-order force constants from the Dynaphopy calculation are used.*

## S6 Ioffe-Regel limit

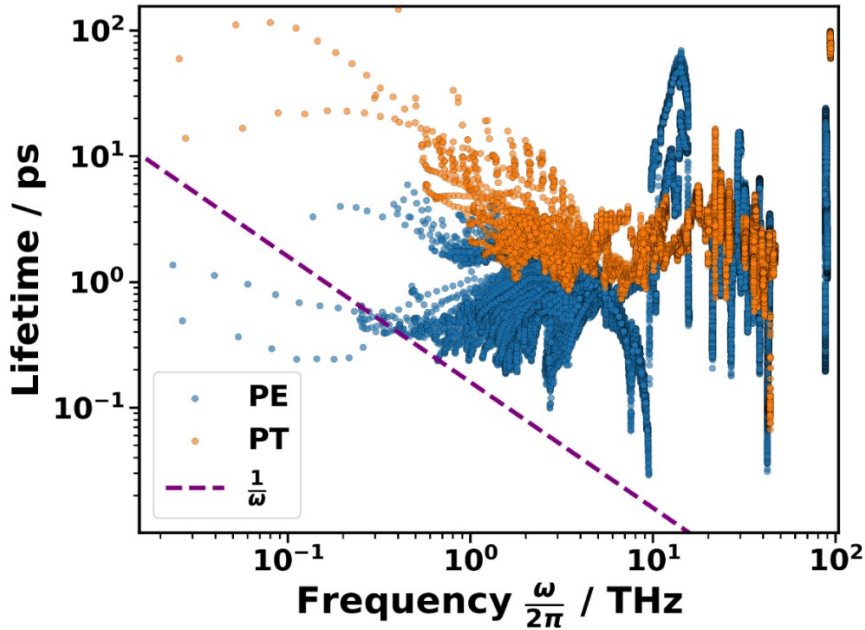

*Figure S14: Phonon lifetimes of PE and PT are plotted against the frequency. The Ioffe-Regel limit is defined by the phonon lifetimes being larger than the reciprocal angular frequency $1/\omega$. The **q**-meshes used in the simulations are 10 × 15 × 160 and 4 × 6 × 48 for PE and PT, respectively.*

As discussed in Ref. [17], for the BTE to be applicable, phonons must fulfil the Ioffe-Regel limit. It states that the lifetimes of phonons need to be larger than their reciprocal angular frequency. The angular frequency is given by $2\pi$ times the "usual" frequency. The latter is used throughout this manuscript, e.g., for plotting the phonon band structure and also in Supplementary Figure S14. As shown in that Figure, the Ioffe-Regel limit is fulfilled for PE and PT, except for a negligible number of phonons of PE.

## S7 Further details on the NEMD calculations

To initialize the NEMD run, we used the following procedure: We first equilibrated the structures at 300 K for 2.5 ps with a Langevin thermostat, then we averaged the energy for 25 ps. We ran MD until a configuration was reached whose energy equaled the average energy within a tolerance of 0.0001%. This configuration was then used as the starting point for the NEMD run. This initialization procedure minimizes effects from temperature oscillations. While we used the Langevin thermostat for the initialization of atomic positions and velocities, in the actual NEMD run, we used the Müller-Plathe method[18] in an *NVE* ensemble employing periodic boundary conditions in all directions and a time step of 0.5 fs. The definition of the regions is the same as in the seminal NEMD paper[18], to which we refer the reader for further details. In short, the NEMD setup consists of a cold region, a region with a temperature gradient, a hot region, and again a region with a temperature gradient, which is in contact with a cold region through the periodic boundary conditions.

The hot/cold region was set to be four unit cells long along the chain direction (for PE and PT). For the calculations along the chain, we found around 15 Å thickness of the simulation box to be sufficient by performing calculations with larger simulation boxes that yielded similar thermal conductivities. The 15 Å thickness of the simulation box corresponds to three and two unit cell repetitions in **y**- and **x**-direction for PE and **x**- and **y**-direction for PT. In the Müller-Plathe method, kinetic energy of the hottest atoms in the cold region are exchanged with the coldest atoms in the hot region, which induces a temperature gradient. This Müller-Plathe energy exchange is implemented via the "fix thermal/conductivity" command in LAMMPS and was performed every 9600 (PE) or every 4800 (PT) time steps for 96 (PE) or 35 (PT) atoms. This low frequency of swaps was found to reduce the total energy drift to a negligible amount. The resulting temperature difference between the hot and cold slab was around 135 K and 180 K for the smallest and largest cell of PT, respectively. For PE, it was 64 K and 104 K for the smallest and largest cell, respectively. The total linear momentum of the hot and cold region was fixed to zero every 100 time steps, which was achieved with the "fix momentum/chunk" command in LAMMPS. This was necessary to cancel any center of mass drift. Simulation times strongly depend on the size of the simulation box, because for larger boxes it takes longer to reach the steady state. For each calculation, we plotted the thermal conductivity over simulation time. The simulations were performed long enough such that the time averaging reduces the statistical noise sufficiently, which is typically the case for a simulation time of 1 ns after the steady state is reached.

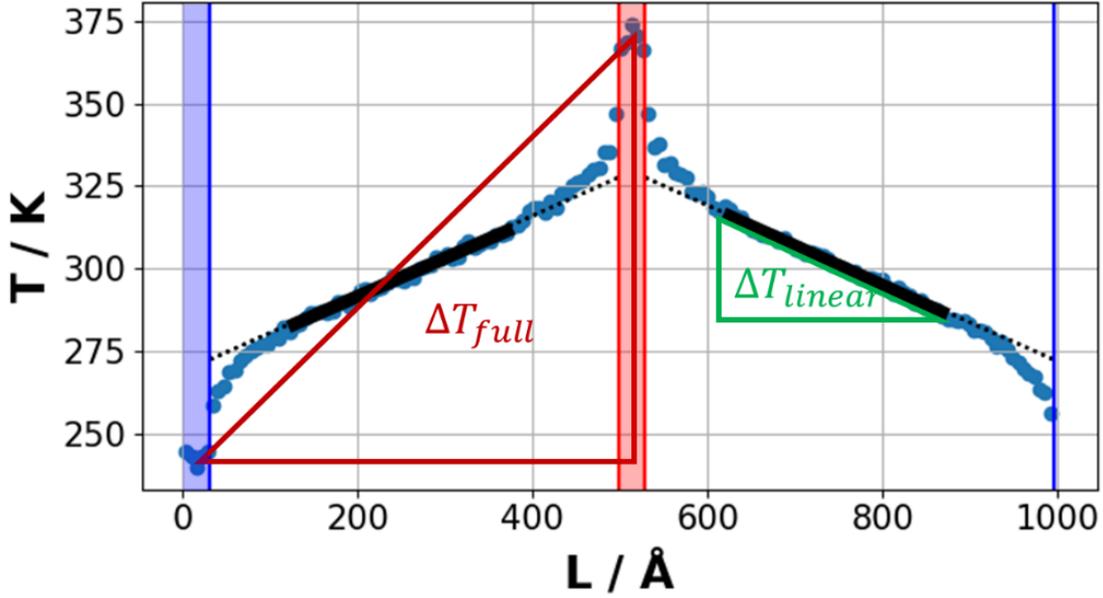

*Figure S15: Typical temperature profile in a NEMD simulation, illustrating the two definitions of the temperature gradient. The cold region spans from 0 Å to 31 Å and is shaded in blue. The hot region is at 498 Å to 529 Å and is shaded in red. The linear part of the temperature profile is illustrated by a black line. Defining the temperature gradient as $\Delta T_{linear}$ takes only the linear part of the temperature profile into account. Li et al. argue that one should rather define the temperature difference as the difference between the thermostat temperatures, leading to the temperature gradient $\Delta T_{full}$.*[19]

In the literature, there exist two different definitions of the temperature gradient in NEMD, which are illustrated in Supplementary Figure S15. According to one definition, one should take only the linear part of the temperature profile, which we refer to as $\Delta T_{linear}$. However, Li et al. found that the nonlinear part should not be excluded, rather one should calculate the temperature gradient from the difference in temperatures between the hot and cold slab.[19] Thus, we calculate this temperature gradient $\Delta T_{full}$ by dividing the temperature difference, which we obtain as the difference between the atom-averaged temperature in the hot and cold thermostat region, by half the box length. For long simulation boxes and small thermal conductivities, the two definitions converge towards each other, since the phonon scattering at the thermostat boundaries becomes small compared to the scattering in the region between the thermostats. This is the so-called diffusive regime. For example, metal organic frameworks (MOFs) have a small thermal conductivity and some of us found that in such a case, the two definitions of the temperature gradient result only in a small difference in thermal conductivity (0.32 Wm$^{-1}$K$^{-1}$ with $\Delta T_{linear}$ and 0.26 Wm$^{-1}$K$^{-1}$ with $\Delta T_{full}$ for MOF-5).[20] However, in our case, we deal with much larger thermal conductivity and, thus, observe a strong relative contribution of the scattering at the thermostat even for the largest boxes that were simulated. Therefore, the definition of the temperature gradient makes a significant difference in our case. The numerical values can be found in Table S11. For PT, NEMD with $\Delta T_{full}$ compares very well to AEMD, MD-BTE, and ALD-BTE, while with the $\Delta T_{linear}$ definition, NEMD significantly deviates from the other methods. Also for PE, the $\Delta T_{full}$ definition yields a better agreement to AEMD and MD-BTE. Therefore, we conclude that one should take $\Delta T_{full}$ as the definition for the temperature gradient, in accordance with Li et al.[19]

*Table S11: Thermal conductivity of PT and PE calculated with different methods. The purpose of this table is to compare the effect that the temperature gradient definition has on the thermal conductivity. The temperature gradient $\Delta T_{full}$ is calculated by taking the temperature difference as the average temperature difference of the thermostat regions, according to the suggestions of Li et al.[19]. $\Delta T_{linear}$ means that the temperature gradient is taken in the linear region. Uncertainties in NEMD and AEMD are the standard deviation (68% confidence interval) taken from the fitting procedure. The values for AEMD, MD-BTE and AEMD are the same as in the main paper. Since for PE in the ALD-BTE calculation, higher order phonon scatterings are relevant (see main paper), the comparison is not particularly helpful here and is thus omitted.*

| Method | $\kappa_{zz}$ of PT / Wm$^{-1}$K$^{-1}$ | $\kappa_{zz}$ of PE / Wm$^{-1}$K$^{-1}$ |
|---|---|---|
| NEMD, $\Delta T_{full}$ | Linear fit: 84 ± 9 | Linear fit: 115 ± 20 |
|  | 2$^{nd}$ order polynomial fit: 98 ± 12 | 2$^{nd}$ order polynomial fit: 146 ± 22 |
| NEMD, $\Delta T_{linear}$ | Linear fit: 131 ± 15 | Linear fit: 197 ± 17 |
|  | 2$^{nd}$ order polynomial fit: 132 ± 27 | 2$^{nd}$ order polynomial fit: 197 ± 47§ |
| AEMD | 94 ± 3 | 127 ± 3 |
| MD-BTE | 98 | 146 |
| ALD-BTE | 91 | - |

§ When using the $\Delta T_{linear}$ definition, the calculation with the smallest unit cell is not converged with respect to the simulation time and is thus omitted in the evaluation of the 2$^{nd}$ order polynomial fit.

Aside of the Müller-Plathe approach, that we used throughout this work, it is also common to employ two thermostats to achieve the temperature gradient. Commonly, local thermostats, such as Langevin thermostats, are preferred. However, when using two Langevin thermostats in our MTP-based simulations of metal organic frameworks for Ref. [20], we encountered large differences between the energy that was added by one thermostat and the energy that was removed by the other thermostat while the total energy in the system was preserved. Thus, we conclude that there must be a technical issue within the energy tallying of the thermostats in LAMMPS (Version 2$^{nd}$ July 2021) when used in combination with the MLIP-LAMMPS interface (Version 2).

## S8 NEMD finite-size extrapolation

To account for the finite box size in the NEMD simulations, one has to perform a finite-size extrapolation, whereby one extrapolates to the bulk thermal conductivity.[21] In this extrapolation, one plots the inverse thermal conductivity over the inverse box size. Then, one typically performs a linear fit and extrapolates to an infinite box size, i.e., inverse box size of zero.[21] This extrapolation can be seen in Supplementary Figure S16 for PE and PT. For PE, the used box sizes had 216, 288, 432, 576, 720, 1080, and 1440 repetitions of the unit cell along the chain direction, while perpendicular to the chains, a fixed number of unit cell repetitions is used as described in Supplementary Section S7. Due to the periodic boundary conditions in chain direction, the distance between the cold and hot thermostat regions is half the box size.[18] The 300 K primitive unit cell of PE is 2.5527 Å long, which result in thermostat distances of 275.69 Å, 367.59 Å, 551.38 Å, 735.18 Å, 918.97 Å, 1378.46 Å, and 1837.94 Å.

Such large supercell lengths are necessary because of the long mean free path of the phonons along the polymer chains.

Since we used vastly different box sizes, one can see non-linear trends in Supplementary Figure S16a, which are expected especially for too small box sizes. Thus, we performed the extrapolations with two approaches: As a first method, a linear fit was made through the datapoints with the largest (three) box sizes. The second method is to fit a polynomial of higher order, in this case a 2$^{nd}$ order polynomial, through datapoints with vastly different box sizes. As shown by Sellan et al.[21], this is justified as a Taylor series expansion of the true but unknown extrapolation function. We argue that both strategies (linear and 2$^{nd}$ order polynomial fit) are sensible, because if one is taking box sizes that are similar in size (e.g., only the largest three box sizes), they typically lie on a straight line and a linear fit works well. If one is trying to perform a fit for boxes of vastly different sizes, they will typically not lie on a straight line, but rather "bend down" (like in Supplementary Figure S16). In that case, it is more sensible to perform a fit of higher order. Sellan et al.[21] noted that such a 2$^{nd}$ order fit is only sensible if multiple data points (at least three but preferably more) are available, which is the case for our PE calculation.

For PE, the aforementioned box sizes give thermal conductivities of 12.55 Wm$^{-1}$K$^{-1}$, 14.96 Wm$^{-1}$K$^{-1}$, 20.27 Wm$^{-1}$K$^{-1}$, 25.76 Wm$^{-1}$K$^{-1}$, 29.65 Wm$^{-1}$K$^{-1}$, 39.75 Wm$^{-1}$K$^{-1}$, and 47 Wm$^{-1}$K$^{-1}$ (using the $\Delta T_{full}$ definition of the temperature gradient, that we discuss in Supplementary Section S7). Extrapolation with a 2$^{nd}$ order polynomial fit gives 145 Wm$^{-1}$K$^{-1}$ ± 22 Wm$^{-1}$K$^{-1}$, where the uncertainty is the standard deviation (68% confidence interval) from the fitting procedure. A linear fit through the largest three box sizes gives 115 Wm$^{-1}$K$^{-1}$ ± 20 Wm$^{-1}$K$^{-1}$. Both values agree with each other within their uncertainties. We note that the extrapolated value is significantly larger than the thermal conductivity obtained for the largest box size, which amounts to 47 Wm$^{-1}$K$^{-1}$. This could be seen as an indication that even larger simulation box sizes would be desirable. However, these go beyond computational capabilities as already now the largest box contains more than 100,000 atoms and for that NEMD simulation the coupled equations of motion need to be solved more than 4 million times.

In analogy to PE, we also performed a finite-size extrapolation for PT. For PT, we calculated boxes with 128, 256, 512, and 1024 unit cell repetitions, which have distances between the cold and hot region of 498 Å, 996 Å, 1992 Å, and 3984 Å. These yield thermal conductivities of 15.14 Wm$^{-1}$K$^{-1}$, 25.16 Wm$^{-1}$K$^{-1}$, 38.10 Wm$^{-1}$K$^{-1}$, and 56.56 Wm$^{-1}$K$^{-1}$. A linear fit through these datapoints gives 84 Wm$^{-1}$K$^{-1}$ ± 9 Wm$^{-1}$K$^{-1}$, while a 2$^{nd}$ order polynomial fit gives 98 Wm$^{-1}$K$^{-1}$ ± 12 Wm$^{-1}$K$^{-1}$ (again, with the $\Delta T_{full}$ definition of the temperature gradient explained below). These data points and the corresponding fits are shown in Supplementary Figure S16b. The situation is somewhat different for PT than for PE, since the linear as well as the 2$^{nd}$ order polynomial fit can fit the datapoints well and also produce rather similar thermal conductivities. This is presumably due to the overall larger supercells considered for this material.

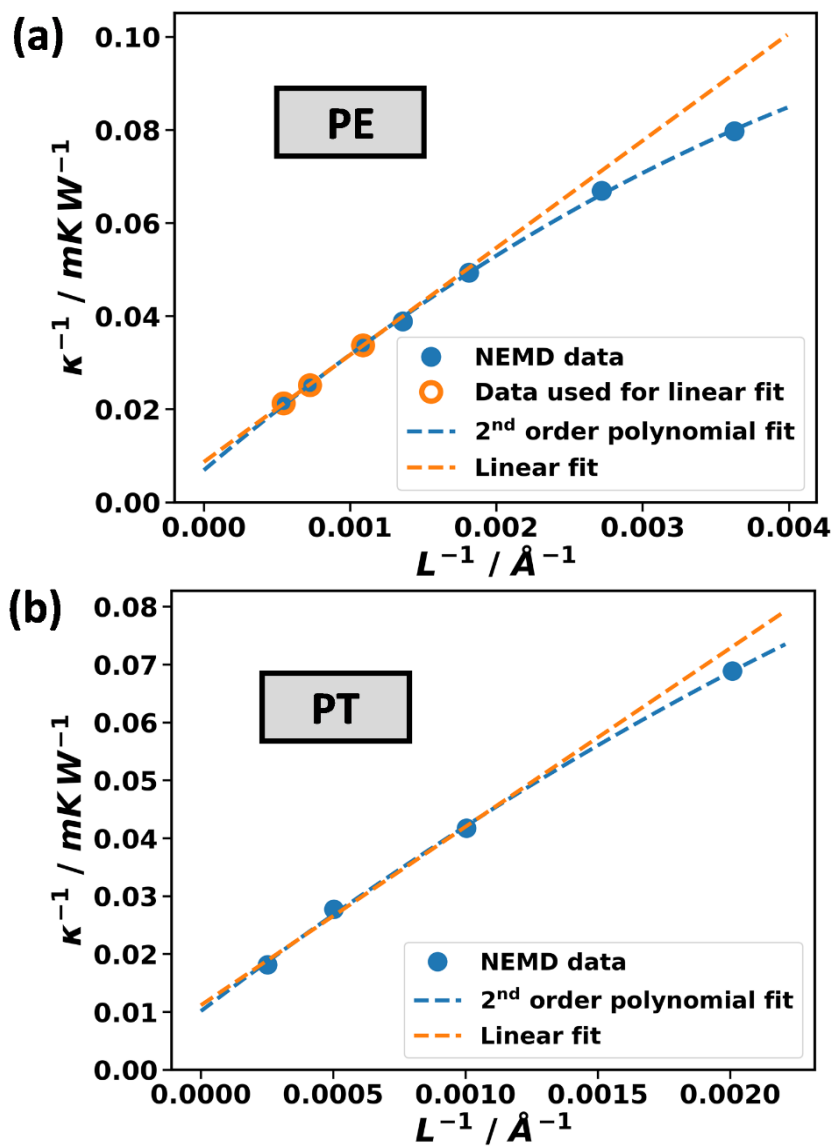

*Figure S16: Finite-size extrapolation of NEMD simulation for a) PE and b) PT. A 2$^{nd}$ order polynomial fit (blue dotted line) and a linear fit (orange line) are performed through the NEMD data points obtained via NEMD simulations on supercells with different lengths, L, along the polymer chains. For PE, the linear fit is only performed through the data points corresponding to the three largest unit cells (highlighted with orange open points). For PT, both fits are performed through all shown data points.*

Still, whether a linear or a 2$^{nd}$ order polynomial fit should be used is unclear and, thus, constitutes a systematic error. The uncertainty from this systematic error is not accounted for in the reported uncertainties, which are solely based on the uncertainty of the respective fits. The inverse thermal conductivities for PT follow the linear fit quite well, even though the box sizes differ by up to a factor of eight. This suggests a rather well-converged extrapolation.

## S9 Further details on the AEMD simulation

The AEMD setup consists of a hot and a cold simulation half. The temperature difference between the hot and cold half $\Delta T(t)$ is calculated from an atomic average over these halves. $\Delta T(t)$ is obtained by a long enough molecular dynamics run (see Supplementary Section S1 for convergence tests) and is subsequently fitted by[22]

$$\Delta T(t) = \sum_{n=1}^{N_{exp}} C_n e^{-\alpha_n^2 \overline{\kappa} t}$$

with the number of exponentials $N_{exp}$, the thermal diffusivity $\overline{\kappa}$ and time $t$. $C_n$ is given by

$$C_n = 8(T_1 - T_2) \frac{\left[\cos\left(\frac{\alpha_n L}{2}\right) - 1\right]^2}{\alpha_n^2 L^2}$$

and

$$\alpha_n = \frac{2\pi n}{L}$$

with the summation index $n$ indicating the order of the exponential, and the length of the cell $L$ (comprising the hot and cold region). Formally, one ought to take $N_{exp}$ as infinite, which is of course not feasible. Luckily, the thermal conductivity quickly converges with respect to $N_{exp}$ as shown in Supplementary Section S1 and in the literature[22]. In our case, we always take $N_{exp}$ equal to 3 as the converged value. From the fit of $\Delta T(t)$, one obtains the thermal diffusivity $\overline{\kappa}$, from which the thermal conductivity $\kappa$ can be calculated via

$$\kappa = \frac{\overline{\kappa} C_p}{V}$$

with the heat capacity at constant pressure $C_p$ and the volume of the supercell $V$. Since we perform classical molecular dynamics (MD) simulations, phonons follow the equipartition theorem[23], and the Dulong-Petit law is valid. The Dulong-Petit law is given by

$$C_p = 3 N_A k_B$$

with the number of atoms $N_A$ and the Boltzmann constant $k_B$.

## S10 AEMD finite-size extrapolation

Similar to the situation for NEMD calculations, also in AEMD one has to correct for finite-size effects by calculating increasingly large supercells and performing a fit to an infinitely large supercell. However, the scatterings are of a different nature than in NEMD. In NEMD, phonons can scatter at the thermostat region, whereas in AEMD there is no thermostat region. Rather, the observation of a length dependence of thermal conductivity in AEMD can be interpreted as a consequence of the extents of the hot and cold regions being smaller than the phonon mean free paths.[24] In the following, we will compare two approaches to perform the finite-size extrapolation in AEMD: a linear fit and a square root fit proposed by Zaoui et al.[24]. The resulting thermal conductivities are denoted by $\kappa_{linear}$ and $\kappa_{Zaoui}$. The linear fit is defined by the function[24]

$$\frac{1}{\kappa(L)} = \frac{1}{\kappa_{linear}}(1 + \frac{\lambda}{L})$$

where $\kappa(L)$ are the thermal conductivity values that we obtain for the different unit cell sizes $L$. $\kappa_{linear}$ is the bulk thermal conductivity that is obtained from the linear fit. In addition to $\kappa_{linear}$, $\lambda$ is also a parameter obtained from the fit with the dimension of a length. The linear fit can be interpreted in terms of a Taylor expansion and could be refined by including more terms[24] (akin to the NEMD fit shown above). However, the linear fit cannot be physically justified in AEMD. Therefore, Zaoui et al. proposed a square root fit of the form[24]

$$\kappa(L) = \kappa_{Zaoui}\left(1 - \sqrt{\frac{\Lambda_0}{L}}\right)$$

where the $\kappa(L)$ are again the thermal conductivities that were obtained at the unit cell sizes $L$. $\kappa_{Zaoui}$ is the bulk thermal conductivity. $\Lambda_0$ as well as $\kappa_{Zaoui}$ are obtained as a result of the fit. In the derivation of the above equation, Zaoui et al. assumed a linear dispersion of the phonon bands at low frequencies up to a frequency of $\omega_0$. The mean free path that corresponds to that frequency is $\Lambda_0 = \Lambda(\omega_0)$. This means that the adjustable parameter $\Lambda_0$ corresponds to the lower limit in the mean free path domain, where the assumption of a linear dispersion is valid. The assumption of a linear dispersion is to some extent a limitation of the Zaoui approach. For PE and PT, this assumption is reasonably well justified, because the acoustic phonons have a roughly linear dispersion and are the main carriers of heat. Zaoui et al. further assumed that the mean free path is proportional to the inverse phonon frequency squared. Since the square root fit is physically motivated, it is preferred here over the more arbitrary linear fit. Table S11 compares the thermal conductivity obtained from AEMD for the two different fit methods. For PE, supercells with 432, 720, 1080, 1440, and 2160 repetitions of the primitive unit cell in chain direction were calculated, which yielded thermal conductivities of 44.3 Wm$^{-1}$K$^{-1}$, 61.1 Wm$^{-1}$K$^{-1}$, 73.0 Wm$^{-1}$K$^{-1}$, 80.6 Wm$^{-1}$K$^{-1}$, and 91.9 Wm$^{-1}$K$^{-1}$ with simulation times given in Supplementary Section S1.4. A square root fit and linear fit give very similar results of 127 Wm$^{-1}$K$^{-1}$ ± 3 Wm$^{-1}$K$^{-1}$ and 125 Wm$^{-1}$K$^{-1}$ ± 3 Wm$^{-1}$K$^{-1}$, respectively (with the standard deviations from the fits as uncertainties). For PT, super cell sizes with repetitions of 128, 192, 256, 384, and 512 were calculated. These give thermal conductivities of 27.2 Wm$^{-1}$K$^{-1}$, 37.1 Wm$^{-1}$K$^{-1}$, 45.8 Wm$^{-1}$K$^{-1}$, 56.6 Wm$^{-1}$K$^{-1}$, and 59.6 Wm$^{-1}$K$^{-1}$. The square root fit yields 94 Wm$^{-1}$K$^{-1}$ ± 3 Wm$^{-1}$K$^{-1}$ and the linear fit 112 Wm$^{-1}$K$^{-1}$ ± 13 Wm$^{-1}$K$^{-1}$. Both fits are consistent with each other within two standard deviations. The linear fit has a much larger uncertainty of 13 Wm$^{-1}$K$^{-1}$, which comes about because the linear fit is not able to fit the data point for the largest cell size well (see Supplementary Figure S17b). Fitting the calculations with large unit cells well is more important than fitting the calculations with small unit cells well. The square root fit performs much better in that regard. For this reason and since it is physically motivated, the square root fit is preferred and, thus, we only give the result for the square root fit in the main paper.

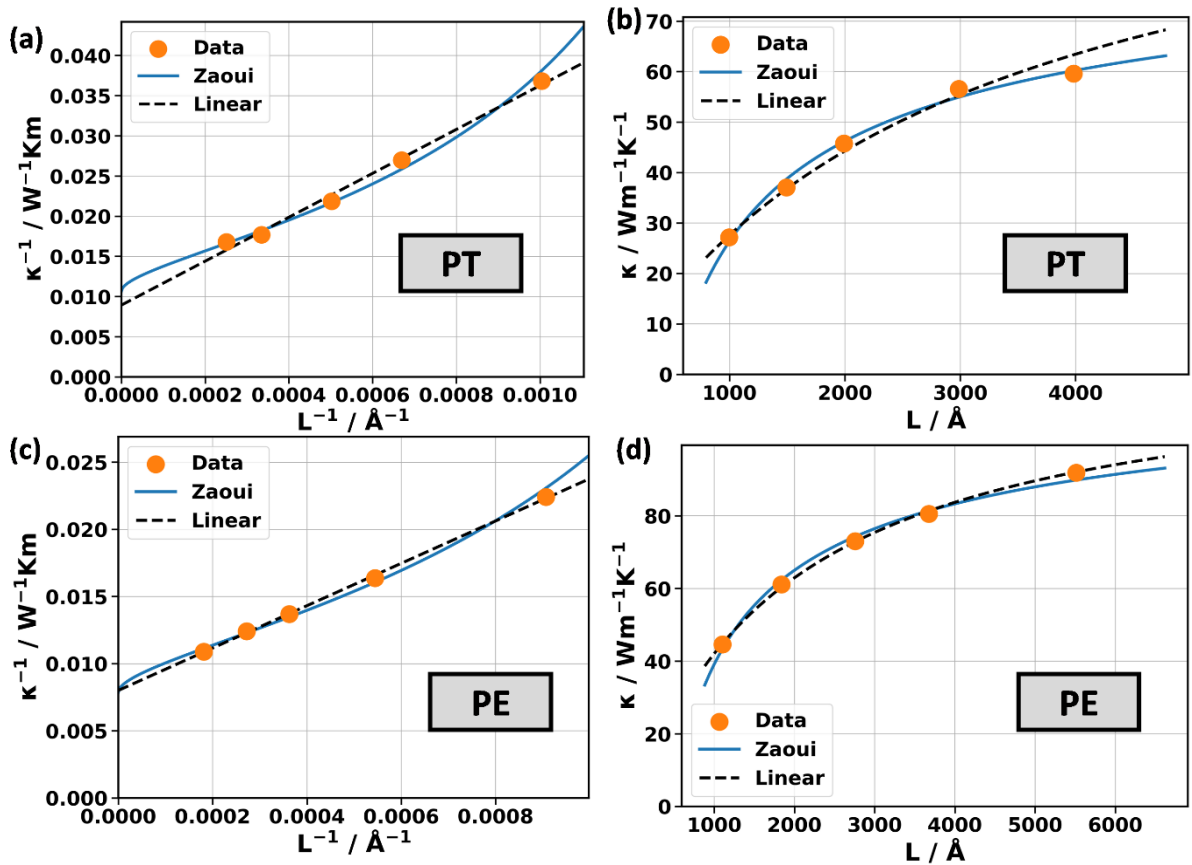

*Figure S17: AEMD finite-size extrapolation for PT plotted for inverse (a) and linear axes (b). (c) and (d) contain equivalent plots for PE. "Zaoui" refers to the square root fit suggested by Zaoui et al.[24], while "Linear" refers to a linear fit of inverse quantities.*

## S11 MTPs for NEMD and AEMD simulations

As discussed in the Methods section of the main paper, MTPs parametrized in a somewhat different way from the procedure suggested in Ref. [1] were used for NEMD and AEMD simulations. When we started working with MTPs, we tested different settings for the MTPs to see which work best. This resulted in MTPs that were trained on a mix of different training data, as described below. These "old" MTPs were used to perform the expensive NEMD and AEMD simulations during the testing phase of the Vienna Scientific Cluster 5 (VSC-5), during which computing hours were not billed. Later, we trained MTPs with more consistent settings, as described in Ref. [1] and in the main paper. Ideally, we would have repeated the NEMD and AEMD simulations with these new "consistent" MTPs. However, this would have been exceedingly expensive, as described in the main paper. Instead of repeating the NEMD and AEMD calculations including the complete finite-size extrapolation, we only repeated the AEMD simulation with relatively small unit cells and show that the "consistent" MTPs give similar thermal conductivities as the "old" MTPs. This is done in Table S12 for PT and Table S13 for PE. To put this comparison into perspective, it is useful to consider that the spread between thermal conductivity values obtained with MTP$^{MD}$s fitted with different initializations is around 8% and 5% for the smallest considered supercell of PT and PE, respectively (see Supplementary Section S16 for details). Here, the difference in thermal conductivity between the "old" MTPs and "consistent" MTPs for the same (smallest) supercell is 8% and 4% for PT and PE, respectively. The extrapolated thermal conductivity varies by 1% and 14% between the two types of MTPs for PT and PE, respectively. Since the difference between the types of MTPs is very similar to the spread of the MTP$^{MD}$s, we argue that the "old" MTPs and "consistent" MTPs can be regarded as essentially equivalent.

Regarding the parametrization of the "old" MTPs: For PT it has a level of 16. Its training data consists of 380 structures sampled with a temperature ramp from 15 K to 500 K and 169 structures sampled at 15 K during an MD run with constant temperature. The "old" MTP of PE is a level 22 MTP with 46 structures sampled at 15 K at constant temperature, 168 structures sampled with a temperature ramp from 15 K to 300 K, and 177 structures sampled at the constant temperature 400 K. For PE, we gave more weight to configurations close to equilibrium by setting the "scale-by-force" keyword to 0.7 in MLIP. As written in the main paper, the "consistent" MTPs are level 22, have training data from 15 K to 500 K and are trained without using the "scale-by-force" keyword. In passing we note that this serves also a test for the level, since the level 16 "old" MTP and the level 22 "consistent data" MTP yield very similar thermal conductivities.

*Table S12: Thermal conductivity of PT calculated with AEMD for different MTPs and different unit cell repetitions along the chain direction. The bulk value is obtained by an extrapolation with the square root fit suggested by Zaoui et al.[24].*

| unit cell repetitions | $\kappa_{ZZ}$ with "old" MTP / Wm$^{-1}$K$^{-1}$ | $\kappa_{ZZ}$ with "consistent" MTP / Wm$^{-1}$K$^{-1}$ |
|---|---|---|
| **128** | 27.2 | 29.3 |
| **192** | 37.1 | 41.7 |
| **256** | 45.8 | 46.8 |
| **bulk** | 89.2 | 90.2 |

*Table S13: Thermal conductivity of PE calculated with AEMD for different MTPs and different unit cell repetitions along the chain direction. The bulk value is obtained by an extrapolation with the square root fit suggested by Zaoui et al.[24].*

| unit cell repetitions | $\kappa_{ZZ}$ with "old" MTP / Wm$^{-1}$K$^{-1}$ | $\kappa_{ZZ}$ with "consistent" MTP / Wm$^{-1}$K$^{-1}$ |
|---|---|---|
| **432** | 43.2 | 44.9 |
| **720** | 61.1 | 62.3 |
| **1080** | 73.0 | 79.6 |
| **bulk** | 120.1 | 136.7 |

# S12 DFT-relaxed and 300 K unit cells

The unit cell length of the DFT-relaxed and 300 K unit cells, that were used for the calculations in the main paper are given in Table S14. The agreement between the DFT- and MTP-relaxed unit cells is good, as shown in much more detail in Ref. [1]. The thermal expansion in vdW-bonded directions is 5% and 1.3% for PE, while it is only 0.9% and 1.0% for PT. This very large thermal expansion of PE leads to significant changes in the thermal conductivity as discussed in the main paper. In stark contrast to the vdW-bonded directions, the thermal expansion in chain direction is very small and negative.

Table S14: Lattice parameters $a_1$, $a_2$, and $a_3$ of the 0 K and 300 K unit cells of PE and PT (see Figure 1 for definitions of the directions). 0 K unit cells are the ones that are relaxed with DFT and the MTP$^{phonon}$ (see Ref. [1] for more details). 300 K unit cells are obtained by performing MD with the MTP (see Methods section of main paper for details).

| Temperature [K] | $a_1$ [Å] | $a_2$ [Å] | $a_3$ [Å] |
|---|---|---|---|
| Polyethylene | | | |
| 0 K, DFT | 7.074 | 4.853 | 2.554 |
| 0 K, MTP$^{phonon}$ | 7.062 | 4.847 | 2.554 |
| 300 K, MTP | 7.445 | 4.918 | 2.553 |
| Polythiophene | | | |
| 0 K, DFT | 7.530 | 5.542 | 7.785 |
| 0 K, MTP$^{phonon}$ | 7.467 | 5.508 | 7.782 |
| 300 K, MTP | 7.600 | 5.600 | 7.780 |

# S13 Complication with the Dynaphopy fit for PT for one of the considered q-points

In Figure 3d of the main paper, the thermal conductivity contribution of the phonons along the shown path are calculated with MD-BTE. The MD-BTE calculations are performed with Dynaphopy, which projects the velocities onto phonon eigenvectors (see Ref. [16] for more details). The result is a power spectrum for each phonon mode and wave vector **q**. Ideally, this power spectrum should contain one Lorentzian peak, that is fitted within Dynaphopy to obtain the phonon lifetime. This is also the case for nearly all considered **q**-points, however, for the longitudinal acoustic phonon at the considered **q**-point closest to the Γ-point, the power spectrum displays two peaks, as shown in Supplementary Figure S18. This is insofar a problem, as it puts the Lorentzian fit for this mode at this wave vector into question. Notably, in the phonopy calculation, there is only a band at around 0.4 THz, but not at 0.38 THz. The shape of this power spectrum also does not change fundamentally, even when the simulation time is increased from 1 ns to 2 ns and the resolution is decreased from 0.004 THz to 0.002 THz. We note that for all other power spectra, that we looked at, there is only one peak. These considerations lead us to the conclusion that the MD-BTE calculation might be less reliable for this particular phonon mode at this particular wave vector but is valid for all other modes and wave vectors. Considering the small volume in reciprocal space that the concerned **q**-point is associated with, the described complication is expected to have no impact on the calculated thermal conductivity.

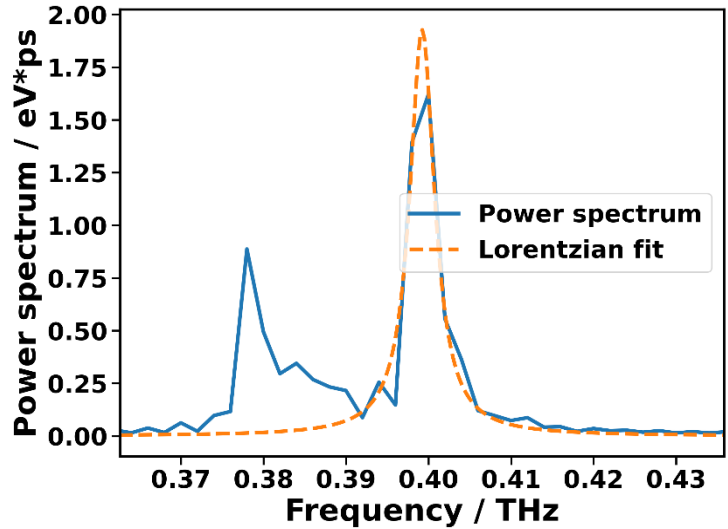

*Figure S18: The Dynaphopy calculation of PT yielded this power spectrum for the longitudinal acoustic mode closest to the Γ-point. The displayed Lorentzian fit was performed with an in-house python script, that takes the lifetime from the Dynaphopy calculation and fits the peak position and peak height. The simulation time is 2 ns and the resolution is 0.002 THz.*

## S14 Frequency-resolved thermal conductivity with heat capacity according to equipartition and Bose-Einstein statistics

To illustrate what difference it makes whether the Bose-Einstein mode heat capacity $C_{BE}$ or the equipartition mode heat capacity $C_{EQ}$ is used, the MD-BTE calculations are evaluated with both heat capacities and the results are shown in Supplementary Figure S19. For PT, where the major contributions to heat transport come from low frequency phonons, the choice of the mode heat capacity has only a small effect. The situation is different for PE, where, due to high frequency phonons, the choice of the heat capacity has a larger effect than for PT.

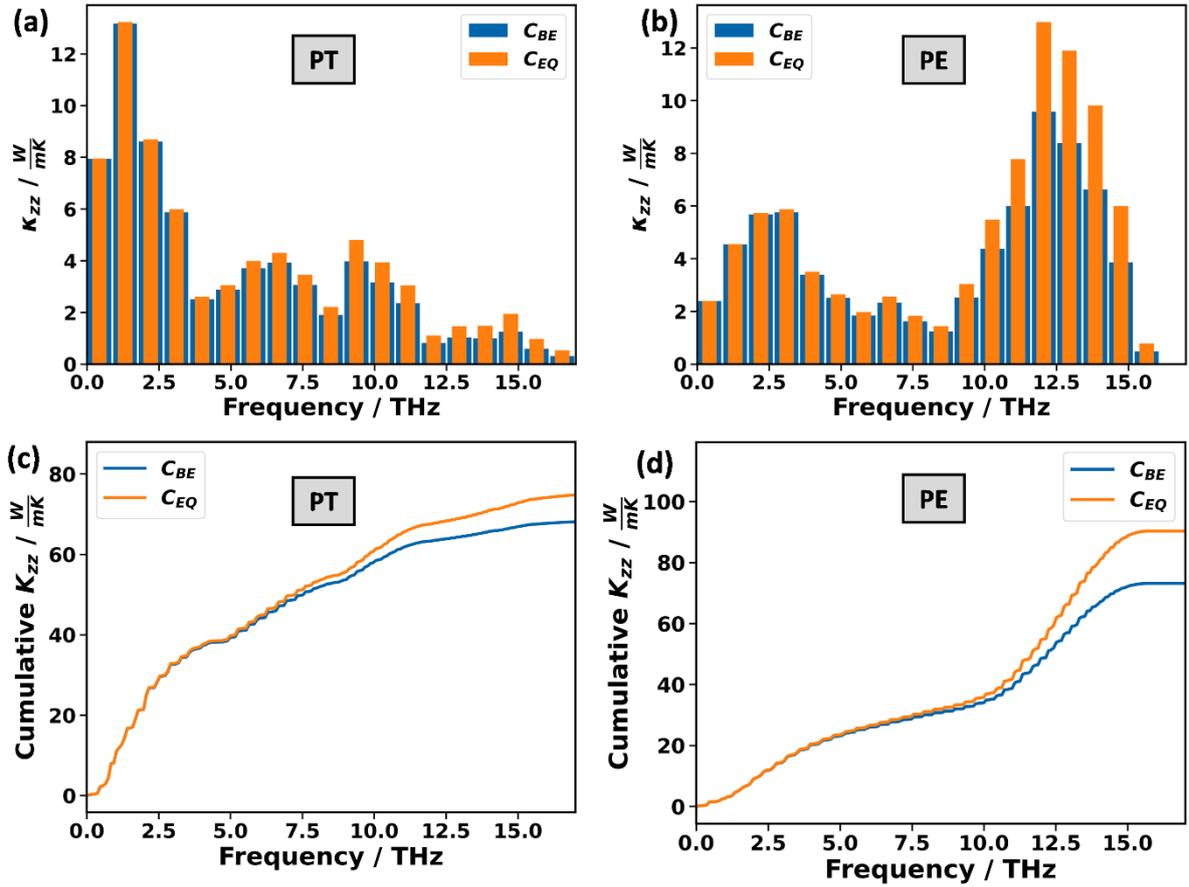

*Figure S19: Frequency-resolved MD-BTE thermal conductivity are calculated with $C_{BE}$ and $C_{EQ}$ assuming a temperature of 300 K for (a) PT and (b) PE. The corresponding cumulative thermal conductivities are shown in panel (c) and (d) for PT and PE, respectively. As for all MD-BTE calculations, the 300 K unit cell is used.*

## S15 Lifetimes of PT calculated with MD-BTE

In the main paper, the lifetimes of PT along the Γ-Z path are given for the ALD-BTE calculation. Here, a respective plot for the MD-BTE calculation is provided in Supplementary Figure S20b. For comparison, the respective ALD-BTE calculation is reprinted from the main paper in Supplementary Figure S20a. This comparison shows that the phonon lifetimes calculated with ALD-BTE and MD-BTE agree rather well with each other. The MD-BTE calculation is performed for 4 ns. Even for this rather long simulation time, the datapoints with the largest lifetimes are somewhat noisy. A calculation with an even longer simulation time could remedy this, but since this would not qualitatively change the result and considering the rather high computational cost, we refrained from performing this calculation.

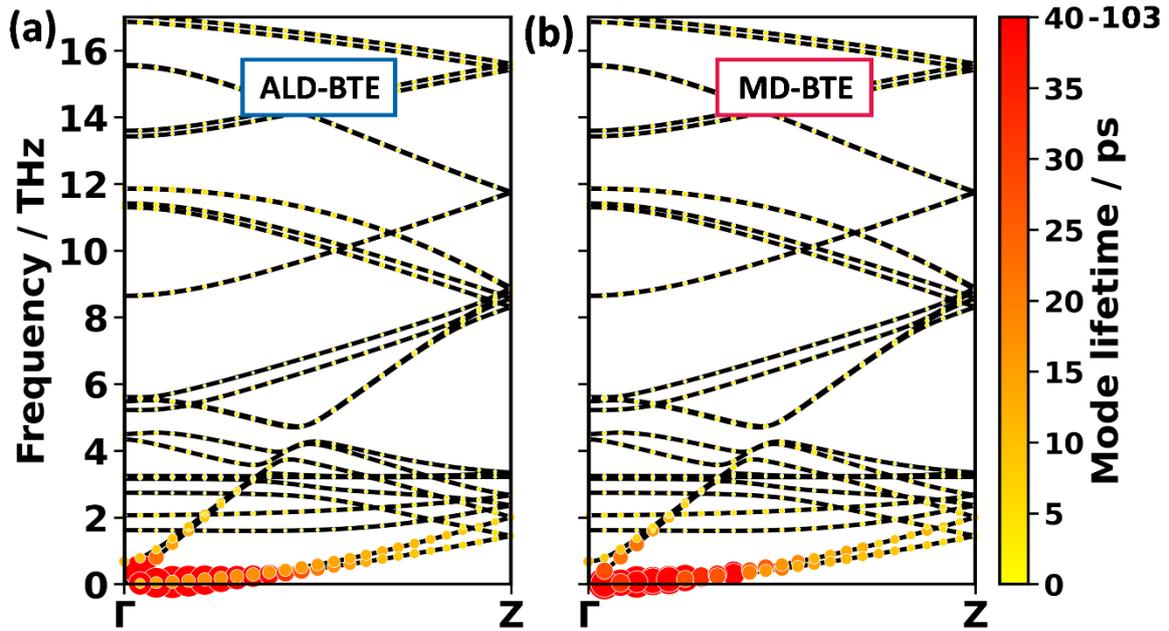

*Figure S20: Lifetimes of PT along the Γ-Z-path calculated with (a) ALD-BTE and (b) MD-BTE. The ALD-BTE plot is the same as in the main paper.*

## S16 Statistical noise in AEMD simulations and uncertainty of MTPs in AEMD

When an AEMD simulation is restarted with a different random seed for initializing the velocities, the resulting thermal conductivity is different due to statistical noise. Typically, we observe that this noise is smaller for larger simulation cells. This is sensible, because the thermal conductivity is effectively evaluated as an "average" over the atoms, and a larger number of atoms thus reduces the statistical noise. Therefore, to estimate an "upper limit" of the statistical noise and to save computational resources, we investigated the noise with the smallest supercell, that was used in AEMD simulations. For PT, this cell contains 21,504 atoms with 128 unit cell repetitions in the chain direction, while for PE it has 31,104 atoms and 432 unit cell repetitions along the chain. Five simulations with different random seeds yield the thermal conductivities listed in Table S15. For both materials, the standard deviation is 5% of the mean value, which we regard as acceptably small. The statistical noise could be reduced by performing each calculation multiple times and taking the average over these calculations. However, since we estimate that the statistical noise is 5% for the smallest supercell and even less for larger cells, we argue that repeating the calculations is not necessary in our case, especially considering the high computational cost that would be associated with repeating the calculations multiple times.

*Table S15: Thermal conductivity with different random seeds for initialization of the velocities in AEMD simulations of the smallest studied supercells. These supercells have the dimensions 2 × 3 × 128 and 2 × 3 × 432 for PT and PE, respectively. Other simulation parameters, including the used MTP, are kept fixed for each material.*

| Index of seed | Thermal conductivity $\kappa_{zz}$ of PT / Wm$^{-1}$K$^{-1}$ | Thermal conductivity $\kappa_{zz}$ of PE / Wm$^{-1}$K$^{-1}$ |
|---|---|---|
| 1 | 25.87 | 44.9 |
| 2 | 28.11 | 42.5 |
| 3 | 29.33 | 41.7 |
| 4 | 29.62 | 47.6 |
| 5 | 29.67 | 46.2 |
| **Mean** | **28.52** | **44.6** |
| **Standard deviation** | **1.44** | **2.4** |

Having determined the statistical noise inherent to the AEMD simulation, we next consider the uncertainty stemming from the MTPs. For this purpose, five MTPs are trained with differently initialized parameters. These MTPs are parametrized according to the MTP$^{MD}$ scheme, as described in the Methods section of the main paper. They yield the thermal conductivities reported in Table S16. For PT, the statistical spread caused by the MTP is around 50% larger than the statistical uncertainty of the differently initialized AEMD simulations discussed above (see Table S15). Still, the overall spread is rather minor considering the small size of the supercells. For PE, the calculated spread due to different MTPs is essentially the same as the statistical noise caused by the AEMD seeds and amounts to around 5%.

*Table S16: Thermal conductivity is calculated using AEMD with five different MTPs for PT (second column) and PE (third column). For PT, the initial seed is set to the seed #3 (see Table S15), therefore the calculation with the third index is identical to the one in Table S15. The supercells have the dimensions 2 × 3 × 128 and 2 × 3 × 432 for PT and PE, respectively.*

| MTP index | Thermal conductivity $\kappa_{zz}$ of PT / Wm$^{-1}$K$^{-1}$ | Thermal conductivity $\kappa_{zz}$ of PE / Wm$^{-1}$K$^{-1}$ |
|---|---|---|
| 1 | 28.99 | 46.2 |
| 2 | 32.27 | 44.9 |
| 3 | 29.33 | 47.1 |
| 4 | 25.77 | 41.5 |
| 5 | 27.40 | 47.5 |
| **Mean** | **28.76** | **45.5** |
| **Standard deviation** | **2.17** | **2.4** |

## S17 Uncertainty of the MTPs in MD-BTE simulations of PE

Considering the complications encountered for the ALD-BTE calculations of PE when using MTP$^{phonon}$s and the 300 K unit cell (see main manuscript), it is worthwhile testing whether similar problems occur for MD-BTE simulations. In fact, in the MD-BTE simulation a much more benign behavior of (in this case MTP$^{MD}$ potentials) was observed: MD-BTE calculations of five distinct MTPs yield 73.3 Wm$^{-1}$K$^{-1}$, 73.4 Wm$^{-1}$K$^{-1}$, 80.3 Wm$^{-1}$K$^{-1}$, 65.9 Wm$^{-1}$K$^{-1}$, and 81.2 Wm$^{-1}$K$^{-1}$. This gives a mean of 74.8 Wm$^{-1}$K$^{-1}$ and a standard deviation of 5.6 Wm$^{-1}$K$^{-1}$, which corresponds to 7% of the mean value. This uncertainty is substantially smaller than for the ALD-BTE simulations, as argued in the main paper. The simulations are performed with a 2 × 3 × 80 supercell, 1 ns simulation time and resolution of 0.004 THz. Here, the supercell is reduced from a 2 × 3 × 160 supercell, as was used in the main paper, to 2 × 3 × 80 to save computational resources, while still employing essentially converged settings.

Supplementary Figure S21 shows that not only the finally obtained thermal conductivities, but also the general trends for the mode contributions to the thermal conductivity are consistent between the different MTPs. In passing we note that based on the smallest force error for the verification set, MTP$^{MD}$ #1 was chosen as the "best" MTP$^{MD}$.

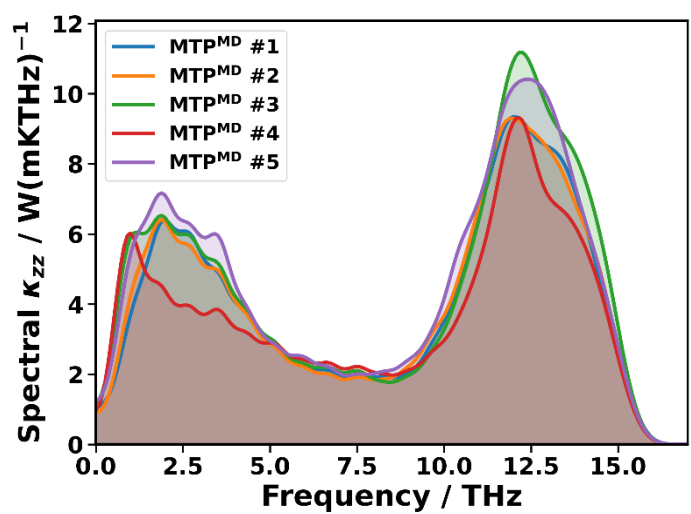

*Figure S21: Spectrally resolved contributions to thermal conductivity of PE calculated with MD-BTE employing five distinct MTP$^{MD}$s. MTP$^{MD}$ #1 is the "best" MTP, that was used throughout the main paper.*

## S18 Thermal conductivities of additional MTPs with ALD-BTE

Typically, we parametrized five MTPs for each configuration (i.e., type of unit cell, material). In the main paper, we reported the "best" out of these and the mean values for the MTP. Here, we report the individual thermal conductivities. The number of negative eigenvalues of the collision matrix is reported in brackets and is discussed in Supplementary Section S1.6.

Table S17: Thermal conductivities of PE in $Wm^{-1}K^{-1}$ for the DFT-relaxed unit cell. The thermal conductivities calculated with five differently initialized $MTP^{phonon}$s are given. MTP #4 is the "best" MTP (see Method Section of main paper for the definition of "best"). The mean value of five differently initialized MTPs is reported with the standard deviation in square brackets. The calculation of the mean for the full BTE calculation is omitted for the 4 × 6 × 120 **q**-mesh because of a negative thermal conductivity. This negative thermal conductivity value is underlined. The number of negative eigenvalues of the collision matrix are given in round brackets (see Supplementary Section S1.6 for more details). The thermal conductivities that are reported in the main paper are highlighted in bold.

|  | MTP index | q-mesh 4 × 6 × 120 | q-mesh 4 × 6 × 160 |
|---|---|---|---|
| RTA | 1 | 229 | 230 |
|  | 2 | 284 | 281 |
|  | 3 | 327 | 324 |
|  | 4, "best" | 291 | **295** |
|  | 5 | 282 | 286 |
|  | mean | 283 [± 35] | **284 [± 34]** |
| full BTE | 1 | 315 (0) | 313 (0) |
|  | 2 | 398 (0) | 385 (1) |
|  | 3 | <u>-159</u> (1) | 430 (1) |
|  | 4, "best" | 435 (2) | **408** (1) |
|  | 5 | 408 (1) | 404 (2) |
|  | mean | - | **388 [± 40]** |

Table S18: Thermal conductivities of PE are reported similarly to Table S17, except that for these calculations the 300 K unit cell is used. Unreasonable thermal conductivities are underlined.

|  | MTP index | q-mesh 4 × 6 × 120 | q-mesh 4 × 6 × 160 | q-mesh 4 × 6 × 320 |
|---|---|---|---|---|
| RTA | 1 | 369 | 363 | 372 |
|  | 2 | 339 | 335 | 341 |
|  | 3, "best" | 265 | **263** | 265 |
|  | 4 | 304 | 304 | 307 |
|  | 5 | 352 | 357 | 359 |
|  | mean | 336 [± 42] | **324 [± 41]** | 329 [± 43] |
| full BTE | 1 | 484 (1) | 491 (2) | <u>642</u> (1) |
|  | 2 | 386 (1) | <u>2527</u> (1) | 506 (2) |
|  | 3, "best" | 303 (1) | **393** (0) | 416 (1) |
|  | 4 | 434 (0) | 325 (1) | 469 (1) |
|  | 5 | <u>884</u> (0) | 415 (2) | 545 (1) |

*Table S19: Thermal conductivities of PT in Wm$^{-1}$K$^{-1}$ for the DFT-relaxed unit cell. The notation is the same as in Table S17.*

|          | MTP index   | q-mesh 4 × 6 × 48 |
|----------|-------------|-------------------|
| RTA      | 1           | 77.0              |
|          | 2, "best"   | **83.9**          |
|          | 3           | 88.0              |
|          | 4           | 73.1              |
|          | 5           | 81.7              |
|          | mean        | **80.8 [± 5.8]**  |
| full BTE | 1           | 93.5 (1)          |
|          | 2, "best"   | **98.4** (0)      |
|          | 3           | 104.8 (1)         |
|          | 4           | 89.1 (0)          |
|          | 5           | 95.9 (0)          |
|          | mean        | **96.3 [± 5.8]**  |

*Table S20: Thermal conductivities of PT in Wm$^{-1}$K$^{-1}$ for the 300 K unit cell. The notation is the same as in Table S17.*

|          | MTP index   | q-mesh 4 × 6 × 48 |
|----------|-------------|-------------------|
| RTA      | 1           | 65.7              |
|          | 2           | 74.3              |
|          | 3           | 72.7              |
|          | 4, "best"   | **76.4**          |
|          | 5           | 71.1              |
|          | mean        | 72.1 [± 4.0]      |
| full BTE | 1           | 79.4 (1)          |
|          | 2           | 90.7 (1)          |
|          | 3           | 84.3 (0)          |
|          | 4, "best"   | **90.5** (1)      |
|          | 5           | 85.5 (0)          |
|          | mean        | 86.1 [± 4.7]      |

# S19 Outlier MTP for the DFT-relaxed unit cell of PE

As listed in the previous section, the MTPs for the DFT unit cell of PE yield 230 Wm$^{-1}$K$^{-1}$, 281 Wm$^{-1}$K$^{-1}$, 324 Wm$^{-1}$K$^{-1}$, 295 Wm$^{-1}$K$^{-1}$ and 286 Wm$^{-1}$K$^{-1}$ in the RTA, generally comparing well to the DFT result of 296 Wm$^{-1}$K$^{-1}$. We regard the MTP that gives 230 Wm$^{-1}$K$^{-1}$ as an outlier, which raises the question whether this outlier MTP yields results that are also qualitatively at variance with the DFT results. In Supplementary Figure S22, the spectral thermal conductivity is plotted for these MTPs alongside the DFT result. The outlier MTP is the one labelled "MTP #1". It is the one that is deviating the most from DFT, but it still reproduces the overall qualitative trend that the largest contributions to the thermal conductivity, when neglecting higher-order scattering processes, stem from phonons between 11 THz and 16 THz.

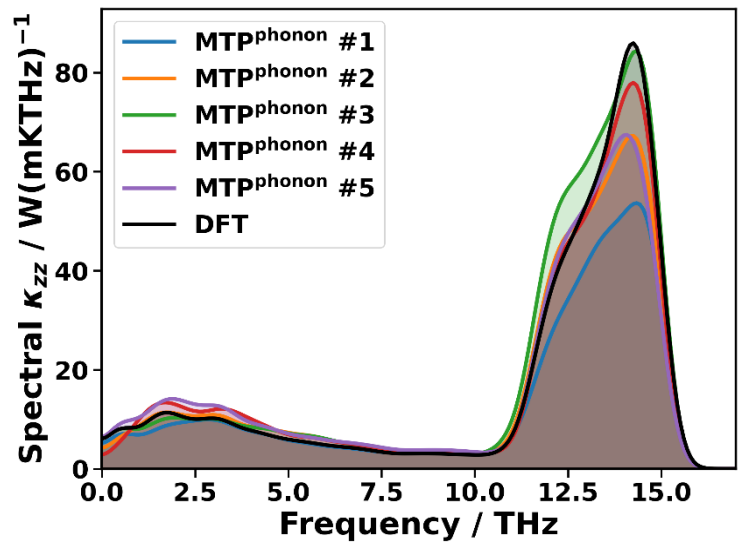

Figure S22: *The spectrally resolved contributions to the thermal conductivity of PE are calculated with the DFT-relaxed unit cell using five MTP$^{phonon}$s (colored lines) and using DFT (black line). "MTP$^{phonon}$ #4" is the "best" MTP.*